\documentstyle[12pt,thmsa,a4,sw20lart]{article}

\input{tcilatex}
\begin{document}

\author{C. Bagnuls\thanks{%
Service de Physique de l'Etat Condens\'{e}e} \ and C. Bervillier\thanks{%
Service de Physique Th\'{e}orique} \\
C.\ E. A.-Saclay, F91191 Gif-sur-Yvette Cedex, France}
\title{Field-Theoretic Techniques in the Study of Critical Phenomena}
\maketitle

\begin{abstract}
We shortly illustrate how the field-theoretic approach to critical phenomena
takes place in the more complete Wilson theory of renormalization and
qualitatively discuss its domain of validity. By the way, we suggest that
the differential renormalization functions (like the $\beta $-function) of
the perturbative scalar theory in four dimensions ($\varphi _{4}^{4}$)
should be Borel summable provided they are calculated within a minimal
subtraction scheme.

\end{abstract}

\section{Introduction}

\hskip\parindent Field-theoretic techniques have appeared efficient in the
obtention of universal properties attached to critical phenomena (for a
review see \cite{zin}) and they are, nowadays, currently used as such.
However these techniques have been questioned by the past (see, e.g., \cite
{par2,nic,nic2,bakkin}) and, recent discussions, relative to some particular
aspects of the approach to criticality, such as the sign of the corrections
to the Ising scaling \cite{liufis,nic3,sokal,3541} or the form of the
classical-to-Ising crossover \cite{3532}, have again put forward questions
relative to their domain of validity in the study of critical phenomena.

By qualitatively describing the relations between the analytical\footnote{%
By analytical we mean a framework which controls a very limited number of
parameters as in perturbation theory.} field theoretic \cite
{gellow,gr,brelegzin} and the Wilson \cite{wilkog} formulations of the
renormalization group (RG) theory, our aim in this article is to illustrate
the origin of the limitations of the former approach. To this end we have
considered the local potential approximation \cite{3480,hashas,fel} of the
renormalization of the scalar theory\footnote{%
Usually called the ``$\varphi ^{4}$-theory''.} that, under the form of a
non-linear partial differential equation, involves the essential properties
of Wilson's exact equations \cite{wilkog,weghou}.

We stress on the fact that this article is only for illustrative purpose,
especially, we do not try to be rigorous (the reader interested in exactness
and/or in explicit mathematical formulas may look, for instance, at \cite
{pol,3094,luswei,sokal,3541}; other recent studies of the local potential
approximation of the RG equations are quoted in \cite{LPA}). The ingredients
of this article are taken from a more detailed version \cite{39}, some of
them have already been the object of short publications \cite
{bagber2,bagber8,inprep2}.

The organization of the article is as follows.

In the following of this part we briefly remind the reader of the main steps
of the analytical version of renormalization with a view to introduce our
notations and some useful definitions. We then briefly present the questions
raised by some authors relatively to the limit of validity of the
field-theoretic approach to critical phenomena \cite{brelegzin}.

In part \ref{SDQFT}, we introduce and formally discuss the notion of
continuous scale dependence in field theory which is the main physical
consequence of the renormalization theory in, say, particle physics\footnote{%
Some authors may consider the renormalization as a simple technique to get
rid of infinities in the perturbation field theory and some others as a
technique to define the continuum limit of lattice field theories.}. We
illustrate how it is used in the field-theoretic approach to critical
phenomena. The object of this part is to bring out some general features of
the analytical framework with a view to recover them in the non-analytic
approach of Wilson considered in part \ref{WEGHOU}. There, we study in some
detail the concrete example of the local potential approximation of the
exact RG equations. We discuss the continuum limit of interest in field
theory with a view to a comparison with the analytical treatment. Especial
emphasis is made on the massless (or critical) theory in three dimensions
and several notions such as the renormalized coupling ``constant'', the ``$%
\beta $-function'' $\cdots $ are presented. We concretely show that a ``$%
\beta $-function'' is associated to the slowest renormalization flow (see 
\cite{wilkog}, p.~132 and \cite{3094}) in the critical (massless)
hypersurface of the Wilson space of Hamiltonian-parameters. We also
concretely show the existence of the slowest renormalization flow
approaching the infra-red stable fixed point from the ``wrong'' side and
responsible for ``negative'' corrections to the Ising scaling \cite{inprep2}%
. Finally, in part \ref{OSAF}, we briefly discuss on the origin of the
singularities encountered in perturbation field theory. We especially argue
that the perturbative series of the $\beta $-function calculated within a
minimal renormalization scheme in four dimensions should be Borel summable.

\subsection{Brief reminder, notations and definitions\label{BRND}}

\hskip\parindent Let us consider the scalar ``Hamiltonian'' involving the
single $\phi _{d}^{4}$ interaction and a momentum cut off $\Lambda _{0}$. 
\begin{equation}
{\cal H}\{\phi \}=\int \!\!d^{d}x\left[ \frac{1}{2}\left\{ (\nabla \phi
(x))^{2}+m_{0}^{2}\phi ^{2}(x)\right\} +\frac{g_{0}\Lambda _{0}^{4-d}}{4!}%
\phi ^{4}(x)\right]  \label{phi4}
\end{equation}

The perturbative renormalization process\footnote{%
Subtraction of infinities that occur at any order of the expansion in powers
of $g_{0}$ of the unrenormalized (or regularized) $N$-point vertex functions 
$\Gamma ^{(N)}({p_{i}};m_{0}^{2},g_{0},\Lambda _{0})$ in the limit $\Lambda
_{0}\rightarrow \infty $ at fixed $m_{0}$ and $g_{0}$ when $d=4$.}
introduces a redefinition of the field ($\phi \rightarrow \phi _{\text{R}}$%
), of the coupling constant ($g_{0}\rightarrow u$) and of the mass ($%
m_{0}\rightarrow m$). Subtraction conditions are necessary to fix the finite
values of the primary divergent parts of the theory. The arbitrariness in
the choice of the finite parts is exploited to give rise to the RG equation 
\cite{gellow,gr,brelegzin} (see Eq. (\ref{RGEsoftmass})).

It is worth to distinguish two main families of subtraction conditions:

\begin{description}
\item[family 1]  {\em ``subtraction point''} schemes that define the finite
parts by explicit conditions on some renormalized vertex-functions
considered at some value of their external momenta (or wave vectors).

\item[family 2]  {\em ``minimal subtraction''} schemes such as obtained by
exclusively subtracting the poles occurring at $d=4$ in the dimensionally
regularized perturbation expansion \cite{dimreg}. They amount to selecting a
peculiar (minimal or the ``slowest'', see section \ref{slowly}) scale
dependence for the renormalized coupling ``constant''.
\end{description}

In the first family, two types of renormalization schemes must be
distinguished:

\begin{description}
\item[type 1]  the massive schemes, in which the limit of a vanishing
renormalized mass ($m\rightarrow 0$) ---i.e., the critical theory--- is not
defined. In such schemes the mass parameter $m$ is similar to the inverse of
the correlation length $\xi $ of statistical systems and provides a
``natural'' scale of reference for the momenta (or the wave vectors).

\item[type 2]  the Weinberg~\cite{wei} schemes, in which the massless theory
is defined. A necessary {\em ``arbitrary''} scale of reference ($\mu $) is
introduced. The renormalized ``mass'' parameter ($t$), when it is different
from zero, is similar to the reduced temperature scale $\left( T-T_{\text{c}%
}\right) /T_{\text{c}}$ of statistical systems undergoing a second order
phase transition.
\end{description}

Family 2 involves only schemes of type 2. For the sake of shortness, from
now on, we shall exclusively consider a renormalization scheme of type 2.

The renormalized vertex (or correlation) functions $\Gamma _{\text{R}%
}^{(N)}(\{p_{i}\};t,u,\mu )$ satisfy the following RG equation \cite
{brelegzin}: 
\begin{equation}
\left[ \mu \frac{\partial }{\partial \mu }+\beta (u)\frac{\partial }{%
\partial u}+(2-\frac{1}{\nu (u)})t\frac{\partial }{\partial t}-\frac{N}{2}%
\eta (u)\right] \Gamma _{\text{R}}^{(N)}\left( \{p_{i}\};t,u,\mu \right) =0
\label{RGEsoftmass}
\end{equation}
in which the differential renormalization functions $\beta (u)$, $\nu (u)$
and $\eta (u)$ are calculable as series in powers of $u$.

The $\beta $-function is of particular interest. It controls the
differential evolution of $u$ as the arbitrary scale $\mu $ is varied: 
\begin{eqnarray}
\mu \frac{du}{d\mu } &=&\beta (u)  \label{beta} \\
\beta (u) &=&-(4-d)u+u^{2}+O(u^{3})  \label{pert}
\end{eqnarray}

\subsubsection{An implicit hypothesis}

\hskip\parindent The perturbative expansion of field theory diverges \cite
{dys1} but the proof \cite{eckmagsen} of Borel summability of the
perturbative series of the {\em massive} $\phi _{3,2}^{4}$ field theories
(in dimension three or two) suggests that the other schemes of
renormalization produce also Borel summable series for $d<4$~\footnote{%
In four dimensions the perturbative series involves ``renormalon''
singularities that do not allow Borel summability (see part \ref{OSAF}).} 
\cite{zin}. In addition to Borel summability, it is however implicitly
assumed that the differential renormalization functions are non-singular at
the nontrivial (infra-red stable) fixed point $u_{\text{ir}}^{*}$, i.e., 
\begin{eqnarray}
\beta (u) &=&\omega (u-u_{\text{ir}}^{*})+O[(u-u_{\text{ir}}^{*})^{2}] \\
\nu (u) &=&\nu +O[u-u_{\text{ir}}^{*}] \\
\eta (u) &=&\eta +O[u-u_{\text{ir}}^{*}]
\end{eqnarray}

\subsubsection{Main ``objections'' encountered in the literature}

\begin{enumerate}
\item  Parisi has objected \cite{par2} that the infra-red singularities
involved in the perturbative expansion of the {\em massless} theory (set 2)
with $d<4$, and which are by-passed by assuming that $\epsilon =4-d$ is an
infinitesimal parameter ($\epsilon $-expansion) \cite{439}, do not allow any
confidence in the estimations of critical exponents in three dimensions from
the massless theory ``{\em {\sl without an additional hypothesis on the
resummation of these infra-red singularities}}'' \cite{par2}. Instead Parisi
advocates the use of the massive framework (set 1) directly in three
dimensions. From another side, it has been shown \cite{sym} that the
elimination of the infra-red singularities of the massless theory ($d<4$)
requires the explicit consideration of parameters, such as the ``critical
bare mass'' $m_{0\text{c}}$ of the dimensionally regularized theory. The
main consequence of this elimination is the appearance of non-perturbative
terms in the expansion \cite{sym}, thus revealing the non-perturbative
nature of field theory \cite{bagber2}.

\item  Baker and Kincaid \cite{bakkin} have emphasized~\footnote{%
However Baker and Kincaid did not make any distinction between the Wilson
and field theoretic versions of the renormalization theory (see \cite
{bagber8}).} that the field-theoretic approach to critical phenomena is
based on the two limits $\Lambda _{0}\rightarrow \infty $ (continuum limit)
and $g_{0}\rightarrow \infty $ (i.e., $u\rightarrow u_{\text{ir}}^{*}$) {\em %
performed in this order}, consequently the framework of field theory could
be unsuited to describe the critical behavior of some systems especially
those which correspond to an infinite $g_{0}$ while $\Lambda _{0}$ is finite
(e.g. the spin-$\frac{1}{2}$ Ising model). A discussion of this
``objection'' has already been presented elsewhere \cite{bagber8} and we
shall not reproduced it in this article.

\item  Liu and Fisher \cite{liufis} and Nickel \cite{nic3} have pinpointed
that the (negative) sign of the corrections to scaling in respectively the
three-dimensional Ising models and the self-avoiding walk models is not
reproduced in the field theoretic framework. This is rigorously true since
the renormalized coupling of the $\phi _{d}^{4}$-theory (with $d<4$) is, by
construction, confined to the range $\left[ 0,u_{\text{ir}}^{*}\right] $ 
\cite{gro} (see also \cite{sokal,3541}) inducing a positive sign for the
corrections to scaling \cite{bagber8,2876,inprep2}. However, the notion of
effective field theory (see section \ref{EFT}) allows us to make sense to
calculations in the range $u>u_{\text{ir}}^{*}$ in the renormalized theory 
\cite{bagber8,inprep2}. The detailed discussion of that point being already
presented in \cite{inprep2}, we shall, again, not reproduce it in this
article (see, however, section \ref{UIASTD}).

\item  Nickel \cite{nic,nic2} has asserted that the $\beta $-function of the
massive scheme in three dimensions involves
confluent-branch-point-singularities at the non-trivial fixed point $u_{%
\text{ir}}^{*}$ implying at least that the quoted uncertainty in the field
theoretic estimates of critical exponents are ``{\em {\sl unrealistically
small}}'' (see part \ref{OSAF}).
\end{enumerate}

\section{Scale-dependence in quantum field theory}

\label{SDQFT}{\sl ``Renormalization is not a technical device to get rid of
infinities but rather is an expression of the variation of the structure of
physical interactions with changes in the scale of the phenomena being
probed.}''~\cite{gro1}

Theories are classified according to the range of length-scales that they
cover: classical mechanics, atomic physics, nuclear physics, particle
physics. That kind of scale-dependence is {\em step-wise}: the physical
parameters involved in a theory are constant in the range of length-scales
that the theory is assumed to cover. The scale-dependence we are presently
interested in differs from the usual notion in that some (or all) physical
parameters of a theory are no longer constant but {\em continuously}\/
depend on the specific length-scale referred to in the range of validity of
the theory. The technique which allows us to define and to deal with such
continuously scale-dependent parameters is called the {\em renormalization
group}\/ first discovered and developed\footnote{%
For technical reasons: the removal of infinities occuring at any order of
perturbation theory.} in studying field theory \cite{gellow,gr}.

The analytical version of renormalization theory yields the following
property for the {\em renormalized}\/ two-point vertex function of the
massless ($t=0$) $\varphi ^{4}$-theory in four dimensions: 
\begin{equation}
\tilde{\Gamma}_{\text{R}}^{(2)}\!\left( \frac{p}{\mu };u\right)
=Z_{3}(\lambda )\tilde{\Gamma}_{\text{R}}^{(2)}\!\left( \frac{p}{\lambda \mu 
};u(\lambda )\right)  \label{udelambda}
\end{equation}
in which we have used $\mu $ to reduce the dimension of the two-point
function $\tilde{\Gamma}_{\text{R}}^{(2)}=\Gamma _{\text{R}}^{(2)}/\mu ^{2}$.

Eq. (\ref{udelambda}) expresses that $\Gamma _{\text{R}}^{(2)}$ takes on
essentially the same form (up to some factor) at external momentum values $%
\tilde{p}$ and $\tilde{p}/\lambda $ (measured in unit of $\mu $) provided
that $u$ and the field $\phi _{\text{R}}(x)$ are changed adequately into $%
u(\lambda )$ and $\phi _{\text{R}}(x)/(Z_{3}(\lambda ))^{1/2}$ respectively.

In Eq. (\ref{udelambda}) the renormalized coupling $u$ is no longer a
constant but a function $u(\lambda )$ for which it is meant that $u$
implicitly depends on the global momentum-scale $\mu $\footnote{$u(1)=u$ is
associated to the momentum-scale $\mu $ while $u(\lambda )$ is associated to
the momentum-scale $\lambda \mu $.}.

In field theory, $u$ may be thought of as a physical coupling strength that
measures the interaction between scalar particles (if any) and it is unusual
to make it continuously depend on a momentum-scale of reference because
usual (i.e. at human scales) physical parameters are fixed. That is why one
usually refers to $u$ as a coupling ``{\em constant}''. In fact, the {\em %
continuous scale dependence of the strengths of interactions}\/ conveys a
fundamental property of field theory~\cite{gellow}. It implies that it is
useless to indicate the strength of an interaction if one does not specify
the momentum (or energy or length) scale which it is attached to. For
example, the usual definition of the electron charge $e$ ($\alpha
=e^{2}/\hbar c=1/137$) by classical macroscopic experiments (like Millikan)
corresponds to very large distances~\cite{shi}.

\subsection{Functional form of the scale dependence\label{FFSD}}

\hskip\parindent If the parameter $u$ depends on $\mu $ continuously, the
analytical version of RG theory does not provide us with {\em the}\/ value
of $u$ which is actually associated to {\em the}\/ given value of the
momentum-scale $\mu $ chosen as reference. The analytical approach to field
theory (perturbative in essence) only provides us with the evolution of $u$
under a change of $\mu $. For example Eq. (\ref{udelambda}) involves $%
u(\lambda )$ (corresponding to the change $\mu \rightarrow \lambda \mu $)
and not the {\em functional form}\/ ``$u(\mu )$''. There is an obvious
technical reason for this fact: {\em $u$ is dimensionless while $\mu $ is
not\thinspace } and in the massless theory no dimensioned parameter other
than $\mu $ is available in perturbation. The main reason, however, is that
the {\em functional form\/ }of the scale dependence is highly
nonperturbative in nature (see parts \ref{WEGHOU} and \ref{OSAF}).

The theoretical $\mu $-dependence of $u$ is, in the analytical treatment,
only written down under the differential form via the $\beta $-function [Eq.
(\ref{beta})]. An initial condition (an external information such as a
measurement) is required to determine the {\em functional form}\/ ``$u(\mu )$%
'' from Eq. (\ref{beta}). When a value $u_{1}$ has been associated to a
given momentum-scale $\mu _{1}$, RG theory allows the determination of all
the other values $u(\lambda )$ associated with the running momentum-scale $%
\mu =\lambda \mu _{1}$: integration of $\beta (u)$ with the given initial
condition, provides the function $u(\frac{\mu }{\mu _{1}})=U\left( \mu /\mu
_{1};u_{1}\right) $. In absence of any initial condition, theory provides us
with a continuous family of functions $U$ indexed by all possible values $%
u_{1}$ that may be associated to a given value of the momentum-scale of
reference $\mu _{1}$~\cite{shi}.

Determining the value of $u$ which is associated to a given value of $\mu $,
we call it determining the {\em functional form}\/ of the momentum-scale
dependence of the coupling strengths. This determination call for
non-perturbative methods (see parts \ref{WEGHOU} and \ref{OSAF}).

\subsection{A technique to derive critical behavior.}

\label{SDUT}\hskip\parindent Let us formally illustrate how the
momentum-scale dependence of the interaction strengths in QFT is used to
study critical behavior.

In order to briefly illustrate the differences which occur in the principles
of using renormalization in field theory in one hand and in the study of
critical phenomena in the other hand, we have drawn two pictures [figs. (\ref
{fig1ter},\ref{fig2})] that qualitatively correspond to the two different
frameworks. Let us discuss those two figures.

\paragraph{Field theory:}

[fig. (\ref{fig1ter})]. Assuming that the ``mass''-parameter $t$ is actually
a measurable quantity, measurements are required in order to fix the
particular values $t_{1}$ and $u_{1}$ which are actually associated to the
arbitrarily chosen momentum scale $\mu _{1}$. This procedure amounts to
selecting an unique (full) curve as indicated on fig. (\ref{fig1ter}).

In absence of an initial condition, there is an infinite family of a priori
allowed curves corresponding to the various ways of associating any allowed
value of $\overline{t}=t(\lambda )/\left( \lambda \mu \right) ^{2}$ (in some
unit $\mu $) to any allowed value of $u$. On fig. (\ref{fig1ter}) we have
only drawn a very small number of such allowed curves (dashed curves).

In field theory with $3\leq d<4$ nothing prevents us from considering the
limit $\mu \rightarrow \infty $ (with $\mu =\lambda \mu _{0}$) corresponding
to an infinite momentum (or energy) scale of reference. It is the
ultra-violet regime the direction of which is indicated by a downward arrow
on fig. (\ref{fig1ter}). In this range of momentum scales, masses are
negligible and thus any (dashed or full) curve reaches the massless
(critical) surface ---the lower axis of fig. (\ref{fig1ter}))--- at the same
point ($\overline{t}=0$ while $u\rightarrow 0$). This limiting point is the
Gaussian fixed point $P_{\text{G}}$ which is ultra-violet stable (for $d<4$%
), we denote it by $u_{\text{uv}}^{*}$. The other fixed point $u_{\text{ir}%
}^{*}$ [set equal to unity on fig. (\ref{fig1ter})] is only reached in the
infra-red limit [small momenta or large distances or low energies, the
direction of which is indicated by an upward arrow on fig. (\ref{fig1ter})]
provided the mass term is kept fixed to zero at any momentum-scale (massless
or critical theory) otherwise the curves go toward the axis $\overline{t}%
=\infty $. We call this latter axis (upper axis of fig. (\ref{fig1ter})) the 
{\em ``trivial surface''}\/ because the physics described in the vicinity of
this ``surface'' corresponds to the classical physics in which the length
scales considered are much larger than the relevant length scales (much
larger than the Compton wavelength ---~or than the correlation length as in
hydrodynamics). On this ``surface'' only classical power laws are expected
whatever the value of $u$. In particle physics this region corresponds to
energies much lower than masses and classical physics is sufficient to
describe it. It must be mentioned here that the ``trivial surface'' may also
be considered as a Gaussian fixed point~\cite{3477} (it is named the
infinite mass or the infinite temperature Gaussian fixed point) although
only one parameter is actually fixed to a given value (the mass which is
infinite).

\paragraph{Statistical physics:}

[fig. (\ref{fig2})]. The use of renormalization theory in studying critical
phenomena is, in nature, very different from the previous description. A
given physical system corresponds there to a given value of $u$, say $u_{0}$%
, associated to a given microscopic length scale $1/\mu _{0}$ (small
compared to the correlation length of the system). The vertical axis at $%
u=u_{0}$ on fig. (\ref{fig2}) intersects the RG flows (full curves) at
points (full circles) corresponding to different values of $\overline{t}$.
These points are representative of the same statistical system ($u_{0}$ is
fixed) observed at different temperatures (measured by reference to the
critical temperature of the system considered). For a definite and
sufficiently small temperature, say $\overline{t}_{0}<1$, a unique full
curve is selected. By following that curve up to $\overline{t}=1$ (up to the
open circle on the selected curve as indicated by black arrows), one
performs an average of the physics occurring at length scale smaller than $%
1/(\lambda \mu _{0})$ (which is larger and larger as $\lambda \rightarrow 0$%
). One thus obtains a description of a new physical system (new values of $u$
and of the field $\varphi _{\text{R}}$). This corresponds to the Wilson step
of thinning out the degrees of freedom of the original statistical system.
The effective system so obtained is no longer the original one, but {\em the
physics}\/ occurring at distances larger than or equal to $1/(\lambda \mu
_{0})$ {\em remains unchanged}\/ compared to the initial system.

By considering smaller and smaller values of $\overline{t}_{0}$, one selects
other full curves on fig. (\ref{fig2}) on which the full circles approach
the critical surface while the open circles approach the axis $u=u_{\text{ir}%
}^{*}$. Hence the limit $\overline{t}_{0}\rightarrow 0$ at fixed $u_{0}$ is
equivalent to the limit $u\rightarrow u_{\text{ir}}^{*}$ at fixed $\overline{%
t}$ provided one is only interested in sufficiently large distances. Notice
that the approach to $u_{\text{ir}}^{*}$ occurs whatever the value of $u_{0}$
and, more importantly, for small enough values of $\overline{t}_{0}$, the
full curves accumulate in the vicinity of the axis $u=u_{\text{ir}}^{*}$
along essentially a unique curve departing from the critical surface $%
\overline{t}=0$. This is a qualitative illustration of universality in
critical phenomena.

\section{A non-perturbative example of renormalization group equations.\label%
{WEGHOU}}

{\sl ``Even if one succeeds in formulating the renormalization group
approach for a particular problem, one is likely to have to carry out a
complicated computer calculation, which makes most theoretical physicists
cringe.''} \cite{wil11}

The main difficulty in understanding the connection between the analytical
and Wilson approaches to renormalization relies upon the number of
parameters involved: a very small number in one hand (essentially the
coupling constant ---renormalized or not) and a very large (infinite) number
in the other hand.

In order to concretely compare Wilson's theory to the analytical field
theory briefly referred to in part \ref{SDQFT}, we consider the differential
RG equations of Wegner and Houghton~\cite{weghou} in the
local-potential-approximation that allows us to reduce the original infinite
set of coupled differential equations of \cite{wilkog,weghou} to a single
partial differential equation (see Eq. (\ref{4})) for a simple function of
the scalar field $\phi $~\cite{3480,hashas}.

\subsection{Brief presentation.}

\label{BPE}\hskip\parindent Let us consider a general Hamiltonian ${\cal H}$
depending on a scalar field $\phi $. It is written as a sum of a Gaussian
part and a potential which, in general, is non-local (i.e. involving
derivatives of $\phi (x)$ that we symbolically denote by $\partial \phi $): 
\begin{equation}
{\cal H}=\int d^{d}x\left[ \frac{1}{2}\left( \partial \phi \right)
^{2}+V(\phi ,\partial \phi )\right]  \label{3}
\end{equation}

The local potential approximation amounts to considering the restriction of
the Wilson space ${\cal S}$ of Hamiltonian-parameters to a smaller space $%
{\cal S}^{\prime }$ such that $V(\phi ,\partial \phi )$ in Eq. (\ref{3})
reduces to its local part $V(\phi )$. Assuming the $O(1)$-symmetry $\phi
\rightarrow -\phi $, we thus have for small $\phi $ : 
\begin{equation}
V(\phi )=\sum_{k=1}^{\infty }a_{k}\phi ^{2k}  \label{potapp}
\end{equation}

It is worth to indicate here that the local potential approximation still
involves essential properties of the ``exact'' RG equations \cite
{wilkog,weghou} since it is similar~\cite{fel} to a continuous version of
both the {\em approximate recursion relation}\/ introduced in \cite{wilkog}
and the {\em hierarchical model}\/ of \cite{dys}.

For the sake of notational simplicity, we denote by $y$ the dimensionless
field $\tilde\phi=\Lambda_0^{(2-d)/2}\phi$ and we shall no longer
distinguish between ${\cal S}$ and ${\cal S}^{\prime}$.

The Wegner-Houghton RG equations in the local potential approximation reduce
to a non-linear partial differential equation~\cite{3480,hashas} for $%
f(y,l)=\partial V(y,l)/\partial y$ with $l$ related to the change of
momentum-scale $\Lambda _{0}\rightarrow \Lambda _{l}=e^{-l}\Lambda _{0}$ (in
the following we also denote the running momentum-scale $\Lambda _{l}$ by $%
\Lambda $). By analogy with an actual flow in usual space, we shall refer to 
$l$ as the ``time'' variable.

Defining $f^{\prime }=\partial f/\partial y$, $f^{\prime \prime }=\partial
^{2}f/\partial y^{2}$, $\dot{f}=\partial f/\partial l$, the partial
differential equation then reads~\cite{hashas}: 
\begin{equation}
\dot{f}=\frac{K_{d}}{2}\frac{f^{\prime \prime }}{1+f^{\prime }}+\left( 1-%
\frac{d}{2}\right) yf^{\prime }+\left( 1+\frac{d}{2}\right) f  \label{4}
\end{equation}
in which $K_{d}$ is the surface of the $d$-dimensional unit sphere divided
by $\left( 2\pi \right) ^{d}$.

Eq. (\ref{4}) for $\dot{f}=0$ is the fixed point equation. Its numerical
study yields the following results. Apart from the trivial Gaussian fixed
point $P_{\text{G}}$ (corresponding to $f\equiv 0$), one observes the
appearance of one new non-trivial fixed point below each dimensional
threshold $d_{k}=2k/(k-1)$, $k=2,3,\ldots ,\infty $. In particular, there is
only one non-trivial fixed point at $d=3$ and none at $d=4$~\cite
{hashas,fel,bagber2}.

\paragraph{Strategy adopted in the present study}

\hskip\parindent We consider Eq. (\ref{4}) at $d=3$, which is the dimension
referred to in the following (except in section \ref{FDC} where $d=4$).

Our aim is to visualize, for the sake of a qualitative comparison with the
analytical field theory, the renormalization flows in the Wilson space $%
{\cal S}$ of Hamiltonian-parameters [the coordinates of which are given by
the coefficients $a_{k}$ of Eq. (\ref{potapp})].

Given an initial simple function, say: 
\begin{equation}
f(y,0)=r_{0}(0)y+u_{0}(0)y^{3}+v_{0}(0)y^{5}  \label{fexp}
\end{equation}
corresponding to a point of coordinates $(r_{0}(0),u_{0}(0),v_{0}(0),0,0,%
\cdots )$ in ${\cal S}$, and after having numerically determined the
associated solution $f(y,l)$ of Eq. (\ref{4}) at a varying ``time'' $l$, we
concretely represent the Wilson trajectories (entirely plunged in ${\cal S}$%
) by numerically evaluating the following derivatives: 
\begin{eqnarray}
r_{0}(l) &=&\left. \frac{\partial f(y,l)}{\partial y}\right| _{y=0}
\label{r_01} \\
u_{0}(l) &=&\left. 6\frac{\partial ^{3}f(y,l)}{\partial y^{3}}\right| _{y=0}
\\
v_{0}(l) &=&\left. 120\frac{\partial ^{5}f(y,l)}{\partial y^{5}}\right|
_{y=0} \\
\text{etc}\ldots &&  \nonumber
\end{eqnarray}

We then are able to visualize the actual Wilson trajectories by means of a
projection onto the planes $\{r_{0},u_{0}\}$ or $\{u_{0},v_{0}\}$ (for
example) of ${\cal S}$.

At $d=3$ we found the non-trivial infra-red stable fixed point ($f^{*}(y)$)
located in ${\cal S}$ at $r_{0}^{*}=-0.461533\cdots $, $u_{0}^{*}=3.27039%
\cdots $, $v_{0}^{*}=14.4005\cdots $, $w_{0}^{*}=32.31289\cdots $, etc. This
is the fixed point to which refers the parameter $u_{\text{ir}}^{*}$ of the
analytical version of RG theory. With a view to illustrate this assertion,
we focus our attention on the massless ---~or critical~--- theory in which
the infra-red stable fixed point is reached when decreasing the momentum
scale of reference down to zero.

\subsection{Determination of the critical surface: the ``shooting method''.}

\hskip\parindent In order to look at the critical renormalization flows, we
must adjust the initial point to be on the critical surface ${\cal S}_{\text{%
c}}$. To this end we use the ``shooting'' method.

As initial function at $l=0$, we consider Eq. (\ref{fexp}) with $v_{0}(0)=0$%
. To start on ${\cal S}_{\text{c}}$ with, say, $u_{0}(0)=3$, we must adjust $%
r_{0}(0)$ [or any other Hamiltonian parameter different from $u_{0}(0)$] to
a non-zero value\footnote{%
Only one hamiltonian-parameter has to be adjusted since ${\cal S}_{\text{c}}$
has the dimension of ${\cal S}$ minus one.}. For any choice of an initial
function at $l=0$ on the critical surface ${\cal S}_{\text{c}}$, there is a
particular value $r_{0\text{c}}\left( 0\right) =r_{0\text{c}%
}(u_{0}(0),v_{0}(0),\cdots )$ of $r_{0}(0)$ for which the nontrivial fixed
point is reached when $l\rightarrow \infty $.

The ``shooting'' method is based on the fact that, for sufficiently large
values of $l$, the renormalization trajectories goes away from $f^{*}$ in
two different directions according to the sign of the difference $r\left(
0\right) -r_{0\text{c}}(0)$.

By numerically determining the values of $r_{0\text{c}}(0)$ corresponding to
different initial points in ${\cal S}_{\text{c}}$, we are able to visualize
the Wilson trajectories approaching $f^{*}$ in ${\cal S}_{c}$.

\subsection{Critical-renormalization-flows and massless field theory in
three dimensions}

\hskip\parindent Some\footnote{%
For not too large values of the initial hamiltonian parameters.}
trajectories in ${\cal S}_{c}$ are represented on fig. (\ref{fig3b}) as a
projection onto the plane $\{u_{0},v_{0}\}$\footnote{%
This choice is arbitrary, we could as well have chosen the plane $%
\{r_{0},u_{0}\}$, for example.}. Let us discuss this figure.

Apart from the fact that each Wilson trajectory goes toward the fixed point $%
f^{*}$ [full circle in the top of fig. (\ref{fig3b})], one observes that
they all approach it asymptotically along the {\em same}\/ one-dimensional
submanifold in ${\cal S}_{c}$. This limiting submanifold, that we denote by $%
T_{1}$, has its source at the Gaussian fixed point $P_{\text{G}}$. Moreover,
the closer the initial points are chosen to $P_{\text{G}}$ [full circle in
the bottom of fig. (\ref{fig3b})], the longer is the ``time'' that the
resulting flows take to go along the {\em unique}\/ submanifold. $T_{1}$ is
an attractive (infra-red stable) submanifold along which the flow are slowly
running (see section \ref{slowly}).

According to Wilson's terminology, $T_{1}$ is a {\em renormalized
``trajectory''}\/ on which is defined the genuine ``continuum limit''%
\footnote{%
In the analytical treatment the limit of interest is that of infinite cutoff
which is called the ``continuum limit'' when the ultra-violet cutoff is
provided by the inverse of a lattice spacing.} of field theory. Hence, to
make sense, the RG flow of the renormalized massless $\phi _{3}^{4}$ field
theory, referred to in part \ref{SDQFT}, must be defined relatively to $T_{1}
$.

Because one practically only encounters in the literature the (sketchy)
description of the massive case as it is presented in section 12.2 of \cite
{wilkog} (i.e., the field theory involving only a mass as renormalized
parameter), we find it worthwhile to examine and discuss the procedure of
``making the cutoff infinite on $T_{1}$'' (the massless case) in some detail.

By analogy with the analytical treatment that refers to a single coupling
``constant'', the procedure consists in showing that some non-zero
``renormalized'' parameter (a coupling ``constant'' for the massless scalar
theory) exists when the initial cutoff tends to infinity.

\subsubsection{The continuum limit of the massless case ($d=3$).}

\label{LICMC}\hskip\parindent The main difference between the well known
limit briefly described in \cite{wilkog} and the case we are considering in $%
{\cal S}_{\text{c}}$ is that the final (``renormalized'') parameter is not a
mass-like parameter (because we are confined to the critical surface) but a
coupling-``constant''-like parameter.

Instead of considering surfaces of constant correlation lengths in ${\cal S}$%
, as in the massive case of \cite{wilkog}, we consider planes of constant $%
u_{0}$ (see section \ref{RUCC}).

Let us select a plane of constant $u_{0}$, say $u_{0}=\overline{u}_{0}=3$,
orthogonal to the $u_{0}$-axis of ${\cal S}$. Let us, then, consider initial
points on ${\cal S}_{\text{c}}$ such that $0<u_{0}(0)<3$ (the other
coordinates being set equal to zero, but $r_{0\text{c}}(0)$ of course). If $%
u_{0}(0)=2$, the corresponding Wilson flow intersects the plane $\overline{u}%
_{0}$ ($=3$) at $Q_{2,3}$ after some finite ``time'' $l_{2,3}$. Let us
denote by $\mu _{0}$ the effective momentum scale of reference associated to
the plane of reference $\overline{u}_{0}=3$ (at present: $\mu
_{0}=e^{-l_{2,3}}\Lambda _{0}$) and let us consider it as the new fixed unit
of momentum\footnote{$\Lambda _{0}$ is then no longer a constant.}. We then
choose another initial point lying closer to $P_{\text{G}}$, say $u_{0}(0)=1$%
. Compared to $Q_{2,3}$, the new intersection-point $Q_{1,3}$ with the plane
of reference is found closer to $T_{1}$ while $l_{1,3}>l_{2,3}$. This means
that, $\mu _{0}$ being fixed\footnote{%
Measured in the previously assumed fixed unit $\Lambda _{0}$, the second
value of $\mu _{0}$ would be smaller than the previous one since $%
e^{-l_{1,3}}\Lambda _{0}<e^{-l_{2,3}}\Lambda _{0}$.}, the initial cutoff $%
\Lambda _{0}$ ``appears to be larger'' at $Q_{1,3}$ than it is at $Q_{2,3}$.
By considering a sequence of initial points $u_{0}(0)=\alpha $ with $\alpha
\rightarrow 0$ (hence approaching the Gaussian fixed point $P_{\text{G}}$)
the sequence of points $Q_{\alpha ,3}$ on the plane of reference $\overline{u%
}_{0}=3$ would finally hit $T_{1}$ at $Q_{0,3}$. At this point, $\mu _{0}$
being kept arbitrarily fixed, the initial cutoff $\Lambda _{0}$ ``appears to
be infinite'' since $l_{\alpha ,3}\rightarrow \infty $ when $\alpha
\rightarrow 0$ (due to the ultra-violet stability of $P_{\text{G}}$). A
non-zero value of a (renormalized) ``$\phi ^{4}$-coupling-constant'' (here $%
\overline{u}_{0}=3$) exists and may thus be associated to an arbitrary
(finite) momentum scale $\mu _{0}$ in the limit of infinite cutoff ($\Lambda
_{0}/\mu _{0}\rightarrow \infty $).

Frequently, one ends the discussion at this point being satisfied by the
existence of the ``infinite-cutoff limit'' involving a {\em single}\/
non-vanishing renormalized coupling ``constant'' (here: $\overline{u}%
_{0}\neq 0$). However the analysis is physically incomplete: what about the
momentum-scale dependence of QFT discussed in part \ref{SDQFT}~? Where is
the effect of the infinite number of degrees of freedom hidden in the limit~?

\subsubsection{Momentum-scale dependence in the continuum limit.}

\label{SDCL}\hskip\parindent A quick inspection of the procedure
just-above-described shows that one may associate any point of $T_{1}$ (but
the ends) to an arbitrary $\mu _{0}$ chosen in the range $]0,\infty [$. It
is usual to say that the massless field theory involves only one {\em %
``free''}\/ parameter (presently, a coupling {\em constant}, $\overline{u}%
_{0}$, the actual value of which lies {\em somewhere}\/ in the range $%
[0,u_{0}^{*}=3.27039\cdots ]$). But no information on the momentum-scale
dependence of $\overline{u}_{0}$ results from these considerations.

The concept of momentum-scale dependence in QFT requires to specify the
succession of points of $T_{1}$ associated with the running momentum-scale $%
\mu =\lambda \mu _{0}$ (with $\lambda $ varying in the range $[0,\infty ]$)
knowing that some given point (say $\overline{u}_{0}=3$) was primarily
associated with the fixed momentum-scale $\mu _{0}$.

After having reached the plane $\overline{u}_{0}=3$, a Wilson flow running
closely along $T_{1}$ continues to go toward $f^{*}$. It intersects the
plane $\overline{u}_{0}=3.2$, for example, at a momentum scale $\mu
_{0}^{\prime }<\mu _{0}$.

\paragraph{Unicity of the differential form: the $\beta $-function.}

For reasons of continuity, both of $T_{1}$ and of the change of
momentum-scale along $T_{1}$, one may easily establish that two points of $%
T_{1}$, chosen infinitesimally close to each other and identified by their
projections (say $\overline{u}_{0}=3$ and $\overline{u}_{0}=3+d\overline{u}%
_{0}$) on the $u_{0}$-axis of ${\cal S}$, are associated to an infinitesimal
change of the momentum-scale of reference $\lambda \mu _{0}\rightarrow
(\lambda +d\lambda )\mu _{0}$ such that: 
\begin{equation}
\lambda \frac{d\overline{u}_{0}}{d\lambda }=f_{1}(\overline{u}_{0})
\label{f1}
\end{equation}
where the function $f_{1}$ is unique for a given {\em representation} of the
renormalized coupling (here a projection of the flow onto the $u_{0}$-axis,
see section \ref{RUCC}). This unicity is illustrated by fig. (\ref{fig4bis}).

The differential expression of the flow along $T_{1}$ depends on the {\em %
representation} of the renormalized coupling. Many different representations
are allowed as discussed in section \ref{RUCC}. But any representation
refers to a {\em unique flow}\/ (``on'' $T_{1}$). Universal features are
thus expected, such as in the asymptotic approach to $u_{0}^{*}$: 
\begin{equation}
f_{1}(\overline{u}_{0})=\omega _{1}(\overline{u}_{0}-u_{0}^{*})+O\!\left[ (%
\overline{u}_{0}-u_{0}^{*})^{2}\right] +O(0)  \label{ff11}
\end{equation}
in which $\omega _{1}$ is a characteristic feature of the Wilson flows along 
$T_{1}$ in the vicinity of $f^{*}$\footnote{%
In the present study, we find $\omega _{1}\simeq 0.6$ (see section \ref
{UIASTD}).}. The term $O(0)$ is present in Eq. (\ref{ff11}) in order to
emphasize that, in principle, the flow of interest runs along the trajectory 
$T_{1}$ which is entirely plunged in a space of infinite dimension. Thus,
one may expect that an infinite number of conditions on the initial
Hamiltonian \cite{sym2}\footnote{%
Symanzik's program for improving lattice field theories (based on
perturbative considerations) relies upon this idea that the limit of
interest in field theory consists in choosing the initial hamiltonian right
on $T_{1}$. Although the number of conditions to be imposed on the lattice
(or unrenormalized) hamiltonian is infinite in principle, it appears to be
finite at each order of perturbation.} should be specified. Actually, the
definition of the differential flow does not require this initial condition
to be specified provided one is able to select directly the particular flow
running along $T_{1}$ (see section \ref{slowly} and part \ref{OSAF}).

Apart the term $O(0)$, the function $f_{1}(\overline{u}_{0})$ is like the $%
\beta $-function of the analytical approach. Only a difference in the choice
of parameterizing $T_{1}$ (via $\overline{u}_{0}$ instead of $u$) occurs
(see section \ref{RUCC}).

\paragraph{Degeneracy of the functional form:}

Contrary to the differential form, the determination of the functional form
of the scale dependence requires to specify explicitly an initial point on $%
T_{1}$ (see section \ref{FFSD}).

Because one cannot specify the infinite set of coordinates of a particular
point lying on $T_{1}$, one arbitrarily chooses an initial point in ${\cal S}%
_{\text{c}}$ in the close {\em vicinity}\/ of $P_{\text{G}}$\footnote{%
The Gaussian fixed point is the only exactly accessible point of $T_{1}$.
But initializing the renormalization process exactly at $P_{\text{G}}$ would
not produce any information on the flow running along $T_{1}$ (by definition
of a fixed point).}. This step corresponds to ``fine-tuning'' the parameters
of the initial Hamiltonian (associated to the momentum-scale $\Lambda _{0}$%
). Due to the freedom in adjusting the initial point near $P_{\text{G}}$, 
{\em an infinite number of definite flows}\/ (Wilson's flows) accumulate
``on'' $T_{1}$. This infinite number indicates the unlimited different
possibilities of associating one point of $T_{1}$ (one value of $\overline{u}%
_{0}$) to a fixed $\mu _{0}$ measured in the unit $\Lambda _{0}$\footnote{%
Due to the infinity of parameters involved in the ``exact'' RG treatment,
the limit of field theory does not actually correspond to the removal of $%
\Lambda _{0}$ which is the seed of the momentum unit.}.

Any ``fine-tuning'' provides us with one determination of the {\em %
functional form}\/ of the momentum-scale dependence in the ``infinite cutoff
limit'' which, thus, is degenerated on $T_{1}$. This is illustrated on fig. (%
\ref{fig04-04bis}).

\subsection{The ``large river effect'' and the notion of ``effective'' field
theory}

\label{FCE}\hskip\parindent A rather close approach to the unique
differential flow numerically determined by reference to the submanifold $%
T_{1}$ occurs independently of whether the initial point is chosen close to
the ultra-violet stable fixed point $P_{\text{G}}$ or not. The convergence
towards $T_{1}$ is exponentially rapid and is sufficient to allow a rather
correct numerical determination of an unique differential flow for a small
enough running momentum-scale of reference.

Indeed $T_{1}$ is like a {\em large river}\/ (having its source ``at'' $P_{%
\text{G}}$) into which watercourses (Wilson's flows) run. (This notion has
recently been used for studying the mechanism of a ''mimicry'' of
second-order phase transitions by fluctuation-induced first-order phase
transitions~\cite{3344}.)

The `large river' effect reveals the existence of a deep valley in ${\cal S}%
_{\text{c}}$ where the renormalization flows are slowly varying (see \cite
{wilkog}, p.~132 and \cite{3094}).

It is interesting to note that the exponentially rapid approach to the
renormalized trajectory is not indicated in the usual (sketchy) graphical
representations of the process of constructing the continuum limit in the
Wilson framework (see, for instance, fig. (1) of \cite{sokal}).

\subsubsection{The flows along the ``renormalized trajectory'' are slowly
varying.}

\label{slowly}\hskip\parindent Let us consider a definite Wilson flow on $%
{\cal S}_{\text{c}}$ not initialized in the close vicinity of $P_{\text{G}}$
but approaching $f^{*}$ asymptotically along $T_{1}$, for example the flow
initialized at $u_{0}(0)=4$, $v_{0}(0)=0$ and $r_{0\text{c}}\left( 0\right)
=-0.381259493\cdots $ [see the corresponding trajectory projected in fig. (%
\ref{fig3b})]. Using obvious notations we shall refer to this flow by $(4,0)$%
.

Such kinds of flow may be decomposed into several parts well separated by
short transitory ranges of ``time''. Disregarding small transitory parts,
each part of a Wilson trajectory is characterized by a regime of
flow-velocities that becomes slower and slower as the part considered is
chosen closer and closer to $T_{1}$. On fig. (\ref{fig3b}) the full curve
corresponding to the trajectory $(4,0)$ clearly displays two regimes%
\footnote{%
Our numerical study of Eq.(\ref{4}) is not accurate enough to correctly
account for the earlier transitory regimes, hence the vertical dashed lines
in the early parts of Wilson's trajectories drawn on fig. (\ref{fig3b}).}.

In order to quantitatively illustrate how the different regimes of
flow-velocities evolve along a Wilson trajectory, it is convenient to
consider the expected analytic form of the effective momentum-scale
dependence of a Wilson flow in ${\cal S}_{\text{c}}$ and in the vicinity of $%
f^{*}$. This gives the following expansions~\cite{weg}: 
\begin{eqnarray}
r_{0}(l) &\stackrel{l\rightarrow \infty }{\simeq }&r_{0}^{*}+b_{r}^{(1,1)}%
\exp \left( -l\omega _{1}\right) +b_{r}^{(1,2)}\exp \left( -2l\omega
_{1}\right) +\cdots  \nonumber \\
&&+b_{r}^{(2,1)}\exp \left( -l\omega _{2}\right) +b_{r}^{(2,2)}\exp \left(
-2l\omega _{2}\right) +\cdots  \nonumber \\
&&+c_{r}^{(1,2)}\exp \left[ -l\left( \omega _{1}+\omega _{2}\right) \right]
+\cdots  \nonumber \\
&&\cdots +\cdots  \label{app1} \\
u_{0}(l) &\stackrel{l\rightarrow \infty }{\simeq }&u_{0}^{*}+b_{u}^{(1,1)}%
\exp \left( -l\omega _{1}\right) +b_{u}^{(1,2)}\exp \left( -2l\omega
_{1}\right) +\cdots  \nonumber \\
&&+b_{u}^{(2,1)}\exp \left( -l\omega _{2}\right) +b_{u}^{(2,2)}\exp \left(
-2l\omega _{2}\right) +\cdots  \nonumber \\
&&+c_{u}^{(1,2)}\exp \left[ -l\left( \omega _{1}+\omega _{2}\right) \right]
+\cdots  \nonumber \\
&&\cdots +\cdots  \label{app2} \\
v_{0}(l) &\stackrel{l\rightarrow \infty }{\simeq }&\text{etc}\cdots 
\nonumber
\end{eqnarray}
in which $\omega _{n}>\omega _{n-1}>\cdots >\omega _{2}>\omega _{1}$.

The smallest exponent $\omega _{1}$ controls the rate of Wilson's flows in
the vicinity of $f^{*}$, it is a characteristic feature of $T_{1}$ in the
vicinity of $f^{*}$. In the field-theoretic approach to critical phenomena 
{\em only one correction-to-scaling exponent, usually denoted by $\omega $,
is available}. In principle $\omega _{1}$ and $\omega $ coincide\footnote{%
Due to the local potential approximation, our estimate of $\omega _{1}$ (see
section \ref{UIASTD}) differs from the usual value of $\omega $ quoted in 
\cite{zin}.}. Any representation of the momentum-scale dependence ``on'' $%
T_{1}$ in the vicinity of $f^{*}$ reduces Eq. (\ref{app1}) (projection onto
the $r_{0}$-axis) and Eq. (\ref{app2}) (projection onto the $u_{0}$-axis) to
their respective first parts, the other exponents ($\omega _{n}$ with $n>1$)
disappear. The correction terms (in the vicinity of $f^{*}$) associated to $%
\omega _{n}$ with $n>1$ are essentially suppressed in Wilson's flows
initialized in the close vicinity of $P_{\text{G}}$ or are extremely
(exponentially) lessened by decreasing the actual momentum-scale of
reference (due to the `large-river' effect). They are merely neglected in
the analytical treatment. When non-exponentially-small compared to the ideal
flow associated to $\omega $, they carry what may be (roughly) called the 
{\em ``finite-cutoff effects''}.

\subsubsection{Notion of effective field theory\label{EFT}}

\hskip\parindent The observation of the ``large river effect'' leads
directly to the notion of ``effective field theories'' (see, for example, in 
\cite{geo}): a reference to the ultra-violet stable fixed point is not
needed to concretely evaluate (within an acceptable accuracy) the behavior
of a field theory below some momentum-scale of reference. This notion is
more significant when no ultra-violet stable fixed point is available (see
sections \ref{OIRASTD} and \ref{FDC}).

\subsubsection{The renormalized coupling ``constant''.}

\label{RUCC}\hskip\parindent In sections \ref{SDCL}, a projection of $T_{1}$
onto the $u_{0}$-axis has been used to refer to the differential form of the
momentum-scale dependence along $T_{1}$ (representation of the
momentum-scale dependence ``on'' $T_{1}$ by means of a single parameter).
The associated scale dependent parameter has been denoted by $\overline{u}%
_{0}$ and called the {\em renormalized parameter}. One must not make a
confusion between $\overline{u}_{0}$ and the {\em unrenormalized}\/ $\phi
^{4}$ coupling constant $u_{0}(0)$. An unrenormalized coupling is attached
to an initial point lying on some {\em canonical}\/ surface of ${\cal S}$
(i.e. it refers to an Hamiltonian involving a limited number of parameters).
The question of dealing with the infinite number of degrees of freedom has
not yet been addressed (a Wilson transformation has not yet been performed).
Instead, in the analytical framework, {\em a renormalized parameter is
defined by reference (under a differential form) to the actual momentum
scale dependence displayed along the attractive submanifold $T_{1}$ (that is
plunged into a space of infinite dimension)}\footnote{%
The correct definition of the renormalized parameter is rather {\em the
relevant parameter at a once infra-red unstable fixed point}\/\/ \cite
{wilkog}. But, in four dimensions the perturbatively renormalized $\phi ^{4}$%
-coupling is (marginally) an irrelevant parameter for the gaussian fixed
point (see section \ref{FDC}), hence the present definition.}.

The choice we have made of referring to $T_{1}$ by a projection onto the $%
u_{0}$-axis could appear arbitrary. In particular, it seems that we could
have chosen any other axis such as $v_{0},w_{0},\cdots $ associated with the
Hamiltonian-terms $\phi ^{6},\phi ^{8},\cdots $~\cite{pol}. This is true as
long as one is only interested in the momentum-scale dependence along $T_{1}$
{\em away}\/ from $P_{\text{G}}$. In this region the slow variation with the
momentum-scale of reference of any running Hamiltonian-parameter is a
characteristic feature of $T_{1}$. In this respect, the $\phi ^{4}$%
-Hamiltonian-term does not distinguish itself from the other
Hamiltonian-terms. We may then choose any Hamiltonian-parameter (or any
combination of Hamiltonian-parameters) to describe the differential form of
the momentum-scale dependence along $T_{1}$. This is not so in the vicinity
of $P_{\text{G}}$ however. There, $T_{1}$ is tangent to the $u_{0}$-axis
(see fig. (\ref{fig3b})) and any other choice of axis is prohibited
otherwise the projection of $T_{1}$ would be empty. The $\phi ^{4}$-coupling
is called the relevant\footnote{%
The relevance of a parameter is an important notion. A relevant parameter
summarizes by itself the conspiracy of infinitely many degrees of freedom
contributing to a field theory.} parameter at the once infra-red unstable
fixed point $P_{\text{G}}$. This is why the $\phi ^{4}$ Hamiltonian-term
plays an essential role in perturbation field theory. Although, farther away
from $P_{\text{G}}$ (in a non-perturbative region) this choice appears
arbitrary, it is the only choice that can describe the momentum-scale
dependence in the whole range of scales from $\infty $ to $0$\footnote{%
The RG is a semi-group (the inverse transform is not defined) the action of
which {\em decreases} the momentum-scale of reference.\label{semi}}.

From now on, the origin of the freedom in the definition of the renormalized
parameter $u$ in perturbation theory (freedom in the choice of the
subtraction scheme) may well be understood~\cite{hashas}. Even in the
perturbative region, the choice of the renormalized parameter is not limited
to $\overline{u}_{0}$. Any vertex function proportional to $\overline{u}_{0}$
in the perturbative region (for example a four-point vertex function
calculated at some value of the set of its external momenta) is an
acceptable choice provided that the differential momentum-scale dependence
along $T_{1}$ be properly accounted for.

\subsection{Other infra-red attractive submanifolds in three dimensions.%
\label{OIRASTD}}

\subsubsection{``Unusual'' infra-red attractive submanifold in three
dimensions.\label{UIASTD}}

\hskip\parindent The trajectories already drawn in fig. (\ref{fig3b}) show
that $T_{1}$, which emerges from $P_{\text{G}}$, is infra-red attractive for
initial points chosen not very far from $P_{\text{G}}$. But, by considering
initial points chosen farther away from $P_{\text{G}}$ in ${\cal S}_{\text{c}%
}$, another infra-red attractive one-dimensional submanifold is evidenced
[see fig. (\ref{fig15})].

Let us call $T_{1}^{\prime }$ this new one-dimensional attractive
submanifold. Which of the two submanifolds $T_{1}$ and $T_{1}^{\prime }$,
corresponds to the slowest flow in the vicinity of $f^{*}$~? Answering this
question requires to considering the expected analytic expression of the
approach to $f^{*}$ in ${\cal S}_{\text{c}}$. It reads: 
\begin{eqnarray}
r_{0}(l) &\stackrel{l\rightarrow \infty }{\simeq }&r_{0}^{*}+b_{r}^{(1)}\exp
\left( -l\omega _{1}\right) +b_{r}^{(2)}\exp \left( -2l\omega _{1}\right)
+\cdots  \label{approach1} \\
u_{0}(l) &\stackrel{l\rightarrow \infty }{\simeq }&u_{0}^{*}+b_{u}^{(1)}\exp
\left( -l\omega _{1}\right) +b_{u}^{(2)}\exp \left( -2l\omega _{1}\right)
+\cdots  \label{approach2} \\
&\cdots &
\end{eqnarray}
with $\omega _{1}>0$. The value of the exponent $\omega _{1}$ is
characteristic of the degree of (infra-red) stability of the submanifold
considered. We choose to denote by $\omega _{1}^{\prime }$ the exponent
associated with $T_{1}^{\prime }$ and by $\omega _{1}$ that associated with $%
T_{1}$. The determination of $\omega _{1}$ (or $\omega _{1}^{\prime }$) from
Eq. (\ref{4}) is made by considering the effective exponent: 
\begin{equation}
\omega _{\text{eff}}=-\frac{1}{r_{0}(l)-r_{0}^{*}}\frac{dr_{0}(l)}{dl}
\label{omegaeff}
\end{equation}
Note that a similar definition of $\omega _{\text{eff}}$ could have been
written down in terms of $u_{0}(l)$ or $v_{0}(l)$ etc.

The limit of Eq. (\ref{omegaeff}) as $l\rightarrow \infty $ gives $\omega
_{1}$ or $\omega _{1}^{\prime }$ according to whether the flow approaches $%
f^{*}$ along $T_{1}$ or along $T_{1}^{\prime }$.

Fig. (\ref{fig11}) reproduces our determination of $\omega _{\text{eff}}$
both along $T_{1}$ and $T_{1}^{\prime }$. It shows that the two flows are
equally slowly running. But the approach to $\omega \sim 0.6$ obtained on $%
T_{1}^{\prime }$ is ``from above'' while on $T_{1}$, it is ``from below''.

By comparison with the usual field-theoretic framework, we are led to
associate the differential momentum-scale dependence of the renormalized
parameter $u$ with the one-dimensional submanifold of ${\cal S}_{\text{c}}$
formed by $T_{1}\cup T_{1}^{\prime }$. The case $u<u_{\text{ir}}^{*}$
corresponds to $T_{1}$ and the case $u>u_{\text{ir}}^{*}$ to $T_{1}^{\prime
} $.

Of course this strict reference to the ``infinite cutoff limit'' defined
right on $T_{1}^{\prime }$ is formal on the field-theoretic point of view
since no ultra-violet stable fixed point exists in association with $%
T_{1}^{\prime }$. Consequently, stricto-sensu, the ``continuum limit'' does
not exist for $u>u_{\text{ir}}^{*}$. Nevertheless, due to the `large river'
effect, the notion of effective field theory applies and gives a meaning 
\cite{inprep2} to (approximated) calculations performed in the analytical
framework for $u>u_{\text{ir}}^{*}$ \cite{inprep2,2876}. The acceptable
accuracy of such calculations from the side $u>u_{\text{ir}}^{*}$ is well
illustrated by fig. (\ref{fig11}) on which we see that the exponent $\omega $
(characteristic of $T_{1}$) may also be determined from the ``wrong'' side
(along $T_{1}^{\prime }$).

The status of the infra-red attractive submanifold $T_{1}^{\prime }$ is
interesting in that several statistical systems ---~especially the spin-$%
\frac{1}{2}$ Ising model in three dimensions~--- correspond to this kind of
approach to $f^{*}$~\cite{bagber8,liufis}.

\subsubsection{Infra-red attractive submanifolds of lower degrees.\label%
{IASLD}}

\hskip\parindent By studying how the approach to $f^{*}$ (along $T_{1}$ or $%
T_{1}^{\prime }$) depends on the initial point chosen in ${\cal S}_{\text{c}%
} $, we find it possible to go continuously from one kind of approach ($%
T_{1} $) to the other ($T_{1}^{\prime }$). This means that, for any given
value of $v_{0}(0)$ (with, say, $w_{0}(0)=\cdots =0$), there exists one
value $u_{0\text{c}}[v_{0}(0)]$\footnote{%
Our $u_{0\text{c}}[v_{0}(0)]$ is identical to the $u_{0\text{c}}[w_{0}(0)]$
introduced on p. 132 of \cite{wilkog}.} for which the resulting trajectory
in ${\cal S}_{\text{c}}$ (i.e. obtained by adjusting also $r_{0}(0)$ to $r_{0%
\text{c}}\left[ u_{0\text{c}}\left\{ v_{0}(0)\right\} ,v_{0}(0)\right] $)
flows toward $f^{*}$ without having any point in common with either $T_{1}$
or $T_{1}^{\prime }$.

Let ${\cal S}_{\text{c}}^{(2)}\subset {\cal S}_{\text{c}}$ be the
hyper-surface which, at $f^{*}$, is orthogonal to $T_{1}\cup T_{1}^{\prime }$%
. ${\cal S}_{\text{c}}^{(2)}$ has one dimension less than ${\cal S}_{\text{c}%
}$. Thus, by adjusting simultaneously two Hamiltonian-parameters in ${\cal S}
$: $r_{0\text{c}}\left( 0\right) $ ---~to be in ${\cal S}_{\text{c}}$~---
and $u_{0\text{c}}(0)$ ---~to be in ${\cal S}_{\text{c}}^{(2)}$~---, we
observe the presence of two new infra-red attractive one-dimensional
submanifolds (or slowly running flows) along which, however, the flows go
faster than along $T_{1}$ and $T_{1}^{\prime }$ (otherwise they would have
been observed first in ${\cal S}_{\text{c}}$).

By studying the Wilson flows in ${\cal S}_{\text{c}}^{(2)}$ we observe the
attractive submanifolds $T_{2}$ and $T_{2}^{\prime }$ drawn in fig. (\ref
{fig16}). As in the preceding case, we have defined an effective exponent $%
\omega _{2\text{eff}}$ and observed that it tends to a unique value~%
\footnote{%
The determination of $\omega_2$ is made less accurate than that of $\omega_1$
because of the need for adjusting two hamiltonian-parameters instead of only
one.} ($\omega _{2}\sim 2.8$) as $l\rightarrow \infty $ both along $T_{2}$
(from below) and $T_{2}^{\prime }$ (from above).

The existence of $T_{2}$ and $T_{2}^{\prime }$ suggests again the existence
of another submanifold ${\cal S}_{\text{c}}^{(3)}\subset {\cal S}_{\text{c}%
}^{(2)}$ in which the approach to $f^{*}$ reveals two infra-red attractive
submanifolds $T_{3}$ and $T_{3}^{\prime }$ with a same exponent $\omega
_{3}>\omega _{2}$ and again one could define ${\cal S}_{\text{c}%
}^{(4)}\subset {\cal S}_{\text{c}}^{(3)}$ and find $\omega _{4}>\omega _{3}$
and so forth$\cdots $

At $f^{*}$, the directions orthogonal to ${\cal S}_{\text{c}}$, ${\cal S}_{%
\text{c}}^{(2)}$, ${\cal S}_{\text{c}}^{(3)}$, ${\cal S}_{\text{c}%
}^{(4)},\cdots $ define a complete set of ``scaling axes'' which the
attractive submanifolds $T_{n}\cup T_{n}^{\prime }$ $(n=0,1,2,\cdots $) are
respectively {\em tangent}\/ to. They correspond to the complete set of
eigenfunctions, solutions of Eq. (\ref{4}) linearized about $f^{*}$. It is
well known that $f^{*}$ possesses one positive eigenvalue corresponding to
the direction of infra-red instability (or relevant direction) orthogonal to 
${\cal S}_{\text{c}}$ (corresponding to the massive sector not considered
here) and infinitely many negative eigenvalues associated with irrelevant
directions (in ${\cal S}_{\text{c}}$, orthogonal respectively to ${\cal S}_{%
\text{c}}^{(2)},{\cal S}_{\text{c}}^{(3)},{\cal S}_{\text{c}}^{(4)},\cdots $%
). These latter directions may be ordered according to the magnitude of the
corresponding eigenvalues (related to the order $\omega <\omega _{2}<\omega
_{3}<\cdots $). The scaling axes have, with respect to $f^{*}$, the same
status as the axes of ${\cal S}$ with respect to the Gaussian fixed point $%
P_{\text{G}}$.

As in the case of $T_{1}$ and $T_{1}^{\prime }$, we may associate, (at least
up to some distance to $f^{*}$) a (formal) ``renormalized'' parameter to
each of those infra-red stable submanifolds $T_{n}$ (or $T_{n}^{\prime }$).
The universal differential momentum-scale dependences (independent of the
initial point chosen in the basin of attraction of the ``renormalized
trajectory'' of interest) associated to the infinite set of submanifolds $%
T_{n}$ ---\thinspace or $T_{n}^{\prime }$\thinspace --- are distinguishable
from each other by their asymptotic rate of flow which may be characterized
by the universal asymptotic-exponents $\omega _{n}$. In the analytical
treatment those new ``renormalized'' parameters are associated to the
renormalization of the insertions of ``composite operators''.

\subsection{The four dimensional case}

\label{FDC}\hskip\parindent The study of Eq. (\ref{4}) at $d=4$ follows the
same line as in $d=3$. The main difference is that the non-trivial fixed
point $f^{*}$ coincides with $P_{\text{G}}$ which becomes (marginally)
infra-red stable in ${\cal S}_{\text{c}}$. Consequently the massless
renormalized trajectory $T_{1}$ (in ${\cal S}_{\text{c}}$) shrinks to a
point ($P_{\text{G}}$) but $T_{1}^{\prime }$ subsists (see fig. (\ref{fig10}%
)). Due to the lack of ultra-violet stable fixed point in ${\cal S}_{\text{c}%
}$, the ``continuum limit'' of field theory does not exist but owing to the
``large river effect'' one may refer to the notion of effective field
theory. The detailed discussion is similar to that of the three dimensional
case for $T_{1}^{\prime }$ and will not be repeated in detail here. The
reader may find supplementary considerations in \cite{hashas,hasnag} for
instance.

\section{The singularities of the perturbation framework\label{OSAF}}

\hskip\parindent Three kinds of singularity are encountered in the
analytical version of scalar field theory:

\begin{enumerate}
\item  the ``renormalons'' that occur in the $\phi _{4}^{4}$-theory \cite
{gronev} and prevent the (renormalized) perturbation series from being Borel
summable.\label{renorm}

\item  the infra-red singularities of the massless $\phi _{d}^{4}$-theory
with $d<4$ \cite{par2,sym}.\label{IR}

\item  the singularities at the infra-red stable fixed point $u_{\text{ir}%
}^{*}$ (for $d<4$) \cite{nic,nic2}.\label{NIC}
\end{enumerate}

The above-listed singularities are peculiar to the analytical treatment.
They emerge from the realization of the following double request:

\begin{description}
\item[A]  the reduction of infinitely many parameters to very few
renormalized (scale dependent) parameters,

\item[B]  the possibility of pushing the momentum scale of reference up to
infinity (existence of an ultra-violet stable fixed point).
\end{description}

These two requirements are explicit (and strictly not dissociated\footnote{%
They are linked together via the unique requirement that the renormalized
parameter is {\em the} relevant parameter at a once infra-red unstable fixed
point.}) in the Wilson process of defining the continuum limit of field
theory (see part \ref{WEGHOU}). In the subtraction program of perturbation
theory, however, both the infinite number of degrees of freedom and the
existence of an ultra-violet stable fixed point are not truly considered
explicitly\footnote{%
The renormalizable character of a theory often appears as the consequence of
a mathematical miracle.}.

In the renormalization program of perturbation theory, the renormalized
parameter is implicitly defined as the parameter having in charge to
exclusively follow the slowest flow in its complete extension (including,
especially implicitly, the infinite momentum scale of reference). Indeed,
the subtraction of the large-cutoff dependences amounts to selecting the
slowest RG flow (only the definition of the renormalized parameter that
exhibits the slowest flow may vary, see section \ref{RUCC}).

The momentum scale dependence of the renormalized scalar coupling involves
two aspects that explain the emergence of the above-listed singularities:

\begin{description}
\begin{itemize}
\item  the functional form of the momentum scale dependence (the account for
infinitely many degrees of freedom)

\item  the domain of variation of the scale of reference ranging from $%
\infty $ to $0$.
\end{itemize}
\end{description}

We have already mentioned that the functional form of the momentum-scale
dependence of $u$ has a non-perturbative nature: a reference to a
supplementary dimensioned parameter $\Lambda _{\text{r}}$\footnote{%
The ``invariant scale'' $\Lambda _{\text{r}}$ is frequently introduced and
used in studies of non-abelian gauge field theory for instance.} is required
to write: $u(\mu /\Lambda _{\text{r}})$. In order to get some idea on the
non-perturbative form expected for $\Lambda _{\text{r}}$ let us integrate
Eq. (\ref{beta}) with the formal assumption that the initial condition is
the existence of the ``fundamental'' invariant-scale $\Lambda _{\text{r}}$.
Hence, the offhand ``definition'' of $\Lambda _{\text{r}}$: 
\begin{equation}
\frac{\mu }{\Lambda _{\text{r}}}=\exp \left[ \int^{u}\frac{dx}{\beta (x)}%
\right]  \label{4.0}
\end{equation}
\noindent in which the undefinite integral stands for the primitive of the
integrand, the additive constant being (by brute force) incorporated in the
definition of $\Lambda _{\text{r}}$. The knowledge of $\Lambda _{\text{r}}$
then would univocally determine the value of $u$ which is actually
associated to $\mu $.

For Eq. (\ref{4.0}) to make some sense, an explicit analytic expression of $%
\beta (x)$ is required. Let us take $\beta (x)$ from the $1/n$-expansion%
\footnote{%
A non-perturbative framework.} with $\epsilon =4-d>0$, Eq. (\ref{4.0}) then
reads: 
\begin{equation}
{\frac{\mu }{\Lambda }}_{\text{r}}=\left| {\frac{\epsilon -u}{u}}\right|
^{1/\epsilon }\left[ 1+O(\frac{1}{n})\right]  \label{4.2}
\end{equation}

Performing the limit $\epsilon \rightarrow 0$ in Eq. (\ref{4.2}), we obtain
at $d=4$: 
\begin{equation}
{\frac{\mu }{\Lambda }}_{\text{r}}=e^{-1/u}\left[ 1+O(\frac{1}{n})\right]
\label{4.3}
\end{equation}

Those (obviously non-perturbative) expressions [Eqs. (\ref{4.2},\ref{4.3})]
are obtained by brute force but they suggest the possible emergence of
pathological effects in perturbation theory. Those effects are the emergence
of the above singularities \ref{IR}.

It is known that the perturbation expansions of vertex functions for the
(super-renormalizable) massless scalar-field theory (with $d<4$) involves
infra-red divergences at rational values of $d$. However, it has been shown 
\cite{sym} that theory ``develops by itself'' an infinite number of
non-perturbative terms that are adapted to make the theory well defined at
any $d$. Among those non-perturbative terms is the ``critical-mass''
parameter $m_{0\text{c}}$. Beside the mathematics, these terms provide us
with nothing but the coordinates (in a space of infinite dimension) of an
(arbitrary) initial point lying on $T_{1}$. Indeed, $m_{0\text{c}}^{2}$ is
similar to the initial critical value $r_{0\text{c}}(0)$ encountered in part 
\ref{WEGHOU}, the difference is that the former implicitly refers to {\em %
some} point of $T_{1}$ (entirely plunged in the Wilson space of infinite
dimension hence the appearance of the other non-perturbative contributions)
while the latter is associated with a chosen initial point of known
coordinates (lying outside from $T_{1}$).

We may say that singularities \ref{IR} reflect the lack, in perturbation
theory, of any specification of an initial point lying on $T_{1}$. It is not
well known that those singularities are also present in four dimensions (for
the massless theory). Indeed, Symanzik has shown \cite{sym} that $m_{0\text{c%
}}$, calculated within the framework of dimensional regularization ($%
\epsilon =4-d>0$), takes on the following form: 
\begin{equation}
m_{0\text{c}}^{2}=g_{0}^{2/\epsilon }h\left( \epsilon ,n\right)   \label{4.4}
\end{equation}
\noindent in which $g_{0}$ is the dimensioned {\em unrenormalized}\/ $\phi
^{4}$-coupling and the function $h\left( \epsilon ,n\right) $ displays poles
at $\epsilon =2/k$ ($k=2,3,4,\ldots $). By considering how the minimal
subtraction scheme of the (dimensionally regularized) $\phi _{4}^{4}$ theory
works within the $1/n$ expansion, Rim and Weisberger \cite{rimwei}
discovered that the subtraction functions of perturbation theory (defined by
the requirement of subtracting the {\em simple}\/ poles located at $\epsilon
=0$ and which introduces the renormalized parameter $u$ associated to the
scale $\mu $) were not sufficient to obtain a well defined theory in the
limit $\epsilon \rightarrow 0$. A new ``mass-counter-term'' (compared to the
perturbative treatment), with a coefficient $\delta m^{2}$ defined as: 
\begin{equation}
\delta m^{2}=\mu ^{2}\tilde{h}\left( \epsilon \right) e^{\frac{2}{u}}
\label{rim}
\end{equation}
\noindent should be added to the dimensionally regularized ``hamiltonian''.
This term arises as a consequence of poles at rational values of $\epsilon $
($\epsilon =2/k$, $k=2,3,\ldots $) {\em which accumulate to give an
essential singularity at $d=4$}. This latter singularity, involved in $%
\tilde{h}\left( \epsilon \right) $, could not be eliminated by the standard
subtraction of simple poles at $\epsilon =0$. Clearly, $\delta m^{2}$ bears
some resemblance with $m_{0\text{c}}^{2}$. Indeed, Rim and Weisberger's
result may be seen as Symanzik's result continued to $\epsilon \rightarrow 0$
at which dimension, the ``infinite-cutoff limit'' of perturbation theory
cannot be effectuated without ``renormalizing the coupling constant''%
\footnote{%
Except the dimension $d$, the difference between Eq.(\ref{rim}) and Eq.(\ref
{4.4}) originates in the renormalization of the coupling constant.}.

In order to show the close relation of these ``additionnal'' terms with the
``invariant scale'' $\Lambda _{\text{r}}$, we have applied the rules of
dimensional renormalization on the critical parameter $r_{0\text{c}}$
calculated at order $1/n$. Denoting by $m_{0\text{c}}^{2}$ the result, we
obtain \cite{bagber2} for $\epsilon >0$ ($d<4$): 
\begin{eqnarray}
\frac{m_{0\text{c}}^{2}}{\mu ^{2}} &=&-\frac{2}{nf(\epsilon )}\left( \frac{%
uf\left( \epsilon \right) }{u_{\text{ir}}^{*}-u}\right) ^{2/\epsilon }\frac{%
\pi }{\sin \left( \frac{2\pi }{\epsilon }\right) }+O\left( \frac{1}{n^{2}}%
\right)   \label{4.7} \\
&\text{with}&u<u_{\text{ir}}^{*}  \nonumber
\end{eqnarray}

\noindent in which $f\,\left( \epsilon \right) =\Gamma \left( d/2\right)
\Gamma \left( 1+\epsilon /2\right) \left[ \Gamma \left( 1-\epsilon /2\right)
\right] ^{2}/\Gamma \left( 2-\epsilon \right) $ ($\Gamma \left( x\right) $
is the Euler gam\-ma-function), and $u_{\text{ir}}^{*}=\epsilon $ in the
approximation considered.

Eq. (\ref{4.7}) is a non-perturbative expression the analytic continuation
to $\epsilon \leq 2$ of which displays poles at $\epsilon =2/k$ ($%
k=1,2,3,\ldots $) which accumulate to give an essential singularity at $%
\epsilon =0$.

In order to allow the limit $\epsilon \rightarrow 0$ to be considered in the
coefficient of the essential singularity of Eq. (\ref{4.7}) it is necessary
to first perform an analytic continuation of Eq. (\ref{4.7}) to $u>u_{\text{%
ir}}^{*}$. Using the relation $(-1)^{2/\epsilon }=\cos \left( \frac{2\pi }{%
\epsilon }\right) +i\sin \left( \frac{2\pi }{\epsilon }\right) $, we obtain: 
\begin{eqnarray}
\frac{m_{0\text{c}}^{2}}{\mu ^{2}} &=&-\frac{2}{nf(\epsilon )}\left( \frac{%
uf\left( \epsilon \right) }{u-u_{\text{ir}}^{*}}\right) ^{2/\epsilon }\pi
\left[ \cot \left( {\frac{2\pi }{\epsilon }}\right) +i\right] +O\left( \frac{%
1}{n^{2}}\right)  \label{riwei} \\
&\text{with}&u>u_{\text{ir}}^{*}  \nonumber
\end{eqnarray}

The limit $\epsilon \rightarrow 0$ performed in the coefficient of $\cot
\left( {\frac{2\pi }{\epsilon }}\right) $ of Eq. (\ref{riwei}) leads to the
``mass-counter-term'' $\frac{1}{2}\delta m^{2}\phi ^{2}$ found by Rim and
Weisberger \cite{rimwei} with: 
\begin{equation}
\delta m^{2}\propto \mu ^{2}e^{\frac{2}{u}}\pi \cot \left( {\frac{2\pi }{%
\epsilon }}\right)  \label{wei4}
\end{equation}

For $\epsilon >0$, Eqs. (\ref{4.7},\ref{riwei}) provide the complete
non-perturbative expression of Symanzik's critical-mass parameter in the
``infinite-cutoff limit'', at order $1/n$ and corresponding to the
submanifolds $T_{1}$ (Eq. (\ref{4.7})) and $T_{1}^{\prime }$ (Eq. (\ref
{riwei})) encountered in part \ref{WEGHOU}.

By comparing the $(u,\mu )$-dependence of Eqs. (\ref{4.7},\ref{riwei},\ref
{wei4}) to that of Eqs. (\ref{4.2},\ref{4.3})), we see that the
non-perturbative properties of $\Lambda _{\text{r}}$ are intimately related
to the non-perturbative properties of the massless field theory. Moreover $%
m_{0\text{c}}$ carries the same $u$-dependence as the ``fundamental''
invariant scale $\Lambda _{\text{r}}$ itself. But the difference is that $%
m_{0\text{c}}$ is absolutely needed (generated by the theory itself \cite
{sym}) while $\Lambda _{\text{r}}$ is a purely formal quantity attached to
the artificial simple function $u\left( \lambda \right) $.

We have seen in part \ref{WEGHOU} that the functional form of the momentum
scale dependence on $T_{1}$ is degenerated (section \ref{SDCL}) and that it
is the specification of the initial value $r_{0\text{c}}(0)$ which allows us
to specify a unique renormalization flow running ``on'' $T_{1}$. Similarly,
the parameter $m_{0\text{c}}$ allows us to specify which determination of
the functional form of the momentum-scale dependence is actually referred to
on $T_{1}$ (or on $T_{1}^{\prime }$).

With the massive theory, the singularities \ref{IR} disappear but the
singularities \ref{renorm} remain. Although the functional form of the scale
dependence is again responsible for the emergence of these singularities,
the mechanism is different from the previous case.

The renormalon singularities may be ``removed'' by considering all
``composite operators'' of dimension 8, 10, etc$\ldots $ \cite{par7,berdav1}
or, what is equivalent, by reintroducing an ultra-violet cutoff \cite
{sym2,berdav1}. It has been shown, for instance, that the perturbative
series in powers of $u_{0}(0)$ of the (unrenormalized) vertex functions of
the $\phi _{4}^{4}$-``infra-red'' model\footnote{%
A massless ``$\phi _{4}^{4}$''-theory with a finite ultra-violet cutoff $%
\Lambda _{0}$ similar to that considered in part \ref{WEGHOU}.} are Borel
summable \cite{felmagrivsen}.

Indeed, the renormalons appear because the trajectory $T_{1}^{\prime }$ does
not take its source at an ultra-violet stable fixed point. Consequently, the
associated domain of momentum scales covered by the running parameter $u$ is
incomplete: when $u$ ``reaches'' infinity, the momentum scale of reference
is still finite. Here is the problem: we do not know how to associate a
genuine momentum scale of reference to a value of $u$ because there are
infinitely many parameters to be initialized. When $u$ is {\em the} relevant
parameter at a fixed point then this association is easy to do: each degree
of freedom delegates one's powers to $u$. But, when there is no ultra-violet
stable fixed point, one must actually specify the complete coordinates of an
initial point lying on $T_{1}^{\prime }$, what is actually impossible. The
other possibility is to consider some Wilson trajectory which will approach
the trajectory $T_{1}^{\prime }$ in reducing the ``cutoff'', this is the $%
\phi _{4}^{4}$-``infra-red'' model mentioned above.

The inverse Borel transforms of the perturbative series of the $\phi
_{4}^{4} $-vertex functions are made ambiguous due to branch-point
singularities located at real positive values of the Borel variable $b$ (the
``renormalon''-singularities). This does not imply that a Borel resummation
of the series cannot be performed but rather that one does not know how to
choose the integration contour in the complex $b$-plane so as to pick up
such and such determination of the integrand at each singularity. Thus an
infinite number of conditions, unknown in the analytic treatment, must be
(re-)specified in order to fix one determination among an infinite number of
allowed determinations (the coordinates of the initial point on $%
T_{1}^{\prime }$ associated to the arbitrarily chosen momentum scale of
reference must be specified).

Notice that, if the above considerations are correct, the emergence of
renormalon singularities is strictly linked to the specification of an
initial point lying on $T_{1}^{\prime }$. If, on a pure mathematical ground
(without any consideration to a well defined field theory), we exclusively
limit our interest to the renormalization flow {\em running} on $%
T_{1}^{\prime }$ via the simple function $u\left( \lambda \right) $, then it
would be possible to imagine a peculiar procedure to construct a $\beta $%
-function the perturbation series of which would be Borel summable.

In our view, this peculiar procedure corresponds to the so-called {\em %
minimal subtraction schemes}. In such a subtraction scheme, nothing else
than the reference to a (specific) scale dependence defines the renormalized
coupling constant $u$, in particular, no reference to a vertex function
(involving the problem of controlling infinitely many degrees of freedom) is
introduced contrary to what is done when using a subtraction point procedure.

The above remarks, that result from the discussions presented in the
preceding parts, illustrate what, indeed, was already expected in four
dimensions. Namely:

{\sl ``that it is possible to define renormalization schemes such that the
renormalization group functions [$\beta (u),\eta (u),\cdots $] do not have
ultra-violet renormalons''\/}\cite{berdav1}.

The arguments presented in this section allow us to propose candidates for
those schemes: the minimal subtraction schemes.

As for the singularities \ref{NIC}, they emerge for the same reason as the
renormalon singularities appear in four dimensions: the lack of ultra-violet
stable fixed point associated with the trajectory $T_{1}^{\prime }$ (i.e.,
in the range $u>u_{\text{ir}}^{*}$).

In view of testing some basical assumptions of the field theoretic approach
to critical phenomena, a lattice-analog of the renormalized $\phi ^{4}$
coupling constant has been defined as follows \cite{bak5}: 
\begin{equation}
u_{\text{latt}}=-\xi ^{-d}\left[ \frac{\partial ^{2}\chi }{\partial H^{2}}%
\right] \chi ^{-2}  \label{glatt}
\end{equation}
in which $\xi $ and $\chi $ are respectively, the correlation length and the
susceptibility of the $\phi ^{4}$-lattice-model; $H$ is the magnetic field.
So defined $u_{\text{latt}}$ ``looks like'' the renormalized $\phi ^{4}$
coupling constant $u$ of the massive scheme in which the renormalized mass $%
m $ is essentially replaced by $\xi ^{-1}$.

By analogy with field theory, the lattice analog of the $\beta $-function is
defined as follows \cite{nic,nic2}: 
\begin{equation}
\beta _{\text{latt}}(u_{\text{latt}})=\left[ -\epsilon \frac{d}{du_{\text{%
latt}}}\ln \left( \tilde{\xi}(u_{\text{latt}})\right) ^{\epsilon }\right]
^{-1}  \label{betalatt}
\end{equation}
in which $\tilde{\xi}=g_{0}^{\frac{1}{\epsilon }}\xi $ with $g_{0}$ the
dimensioned unrenormalized $\phi ^{4}$ coupling constant and $\epsilon =4-d$.

Nickel's discovery of confluent singularities in the $\beta $-function
follows from the remark that renormalization theory predicts \cite{weg}
several kinds of confluent critical singularities. The approach of $u_{\text{%
latt}}$ to $u_{\text{ir}}^{*}$ as $\tilde{\xi}\rightarrow \infty $ is, thus,
controlled by the following equation: 
\begin{eqnarray}
u_{\text{latt}} &=&u_{\text{ir}}^{*}\left[ 1+a_{1}\tilde{\xi}^{-\omega
_{1}}+a_{2}\tilde{\xi}^{-2\omega _{1}}+\cdots \right.  \nonumber \\
&&+b_{1}\tilde{\xi}^{-\omega _{2}}+b_{2}\tilde{\xi}^{-2\omega _{2}}+\cdots 
\nonumber \\
&&\left. +\cdots +c_{1}\tilde{\xi}^{-\frac{1}{\nu }}+c_{2}\tilde{\xi}^{-%
\frac{2}{\nu }}+\cdots \right]  \label{tau}
\end{eqnarray}
in which the exponents $\omega _{n}$ have already been encountered in
sections \ref{FCE} and \ref{IASLD}. Supplementary terms proportional to $%
c_{n}$ account for analytic confluent corrections in temperature usually not
explicitly considered by renormalization theory and thus not relevant to the
present discussion.

Now, on reverting Eq.(\ref{tau}), and neglecting the above-mentioned
analytic confluent corrections in temperature, one obtains: 
\begin{eqnarray}
\tilde{\xi} &\simeq &a_{0}(u_{\text{ir}}^{*}-u_{\text{latt}})^{-1/\omega
_{1}}\left[ 1+a_{2}^{\prime }(u^{*}_{\text{ir}}-u_{\text{latt}})+\cdots
\right.  \nonumber \\
&&\left. +b_{1}^{\prime }(u^{*}_{\text{ir}}-u_{\text{latt}})^{(\omega
_{2}-\omega _{1})/\omega _{1}}+\cdots \right]  \label{xilatt}
\end{eqnarray}
and finally, using Eq.(\ref{betalatt}): 
\begin{eqnarray}
\beta _{\text{latt}}(u_{\text{latt}}) &\simeq &\omega _{1}(u_{\text{latt}%
}-u^{*}_{\text{ir}})+d_{1}(u^{*}_{\text{ir}}-u_{\text{latt}})^{2}+\cdots 
\nonumber \\
&&+d_{2}(u^{*}_{\text{ir}}-u_{\text{latt}})^{\Delta _{2}/\Delta _{1}}+\cdots
\label{betalatt2}
\end{eqnarray}
in which $\Delta _{n}=\omega _{n}\nu $.

Obviously, the second line of Eq. (\ref{betalatt2}) displays a leading
confluent-branch-point-singularity that is not accounted for in the
resummation procedure of $\beta (u)$ as used in \cite{legzin}. Nickel then
makes reference to the work of Golner and Riedel \cite{golrie} who found $%
\Delta _{2}\simeq 2\Delta _{1}$ (see also the more recent work of Newman and
Riedel \cite{newrie} that yields in addition $\Delta _{3}\simeq 3\Delta _{1}$%
) to argue that the confluent singularities could be practically very weak
as explicitly assumed later on by Zinn-Justin \cite{zin7}. If so, Nickel
does not understand

{\sl ``$\cdots$ why the confluent terms in the lattice-analog $\beta_{\text{%
latt}}(u_{\text{latt}})$ can apparently not be ignored''}

\noindent in an actual numerical analysis of high-temperature series.

At the light of the preceding parts, that may well be understood.

The confluent singularities found by Nickel in $\beta _{\text{latt}}$ are
due to what we have called the ``finite-cutoff effects'' (see section \ref
{slowly}). These effects ``measure'' the significant deviations, in the
critical surface ${\cal S}_{\text{c}}$, between a simple Wilson trajectory
(of a lattice or ``cutoff'' theory) and the ideal ``trajectory'' $T_{1}$ (or 
$T_{1}^{\prime }$) of part \ref{WEGHOU}. Consequently, in the field
theoretic framework, Nickel's confluent singularities within the $\beta $%
-function are either absent ($u<u_{\text{ir}}^{*}$, by definition of the
``continuum limit'') or are extremely weak ($u>u_{\text{ir}}^{*}$) due to
the `large river' effect\footnote{%
The fact that one deals with the massive theory does not matter since we are
more interested in the vicinity of $u_{\text{ir}}^{*}$ where mass effects
vanish.}.

The weakness of the confluent singularities could explain why, in their
careful studies of the various available series, Le~Guillou and Zinn-Justin 
\cite{legzin} have not observed any numerical effect which could be clearly
associated to the existence of such singularities.

Despite the supposed weakness of the confluent singularities in the range $%
u\geq u_{\text{ir}}^{*}$ for the massive framework in three dimensions,
attempts at systematically accounting for them have been made \cite{nic2}
(on the basis of the knowledge of seven orders instead of six orders only,
except for the function $\beta (g)$ for which only the already known six
orders were used). Their study has provided new exponent estimates which 
{\sl ``are in better agreement with other model calculations''}. This result
could be seen as a concrete confirmation of the presence of confluent
singularities in the range $u\geq u_{\text{ir}}^{*}$. But, due to the lack
of relevant information on the actual value of $\Delta _{2}$ and on higher
order terms in the series of $\beta (u)$, it has only been concluded that
the previous error bars \cite{legzin} could be {\em ``unrealistically small''%
}. It is also fair to indicate that, due to the lack of order in the series
of $\beta (u)$ a bias could have been introduced in the determination of the
fixed point value $u_{\text{ir}}^{*}$.

\newpage

\begin{center}
{\bf FIGURE CAPTIONS}
\end{center}

\begin{enumerate}
\item  Qualitative representation of the momentum-scale dependence for the $%
\phi _{d}^{4}$ field theory with $d<4$. The curves drawn (see also figure 
\ref{fig2}) are obtained from assuming $\nu =1/2$ and a $\beta $-function of
the form $\beta (x)=x(x-1)$; $\overline{t(\lambda )}$ stands for $\tilde{t}%
(\lambda )/\lambda ^{2}$. Among the continuous family of curves (dashed
curves) a measurement is required in order to choose (full circle) one curve
(full curve). An upward arrow indicates the infra-red direction, a downward
arrow the ultra-violet direction. See text for more detail. \label{fig1ter}

\item  Use of the $\phi _{d}^{4}$ field theory in the study of critical
phenomena. The presentation is similar to fig.(\ref{fig1ter}). See text for
a discussion of the figure. \label{fig2}

\item  Projection onto the plane $(u_{0},v_{0})$ of some Wilson's
trajectories on the critical surface ${\cal S}_{\text{c}}$ (after
integration of Eq.(\ref{4}) at $d=3$). Full circles indicate the locations
of the Gaussian ($P_{\text{G}}$) and the non-trivial ($f^{*}$) fixed points
respectively. Arrows indicate the infra-red direction. Open circles indicate
the locations of the initial points. The values of the initial sets of
coordinates $(r_{0\text{c}}(0),u_{0}(0),v_{0}(0))$ are the following: ($-$%
0.014175577$\cdots $, 0.1, 0.), ($-$0.116887403$\cdots $, 1., 0.), ($-$%
0.299586913$\cdots $, 3., 0.), ($-$0.381259493$\cdots $, 4., 0.), ($-$%
0.180075418$\cdots $, 1., 5.), ($-$0.025823782$\cdots $, 0., 1.) the other
coordinates being set equal to zero. The full curves represent Wilson's
trajectories entirely plunged in ${\cal S}$. The ``vertical pieces''
(dot-dashed lines), are numerically undetermined parts of the projected
Wilson trajectories. They reflect the lack of accuracy in our determination
of $v_{0}(l)$ for small $l$. \label{fig3b}

\item  Graphical representation of the Wilson differential flows (dashed
curves), in ${\cal S}_{\text{c}}$ at $d=3$, projected on the $u_{0}$-axis
--i.e. $f_{1}(\overline{u}_{0})=-d\overline{u}_{0}(l)/dl$. In the infra-red
direction, any (dashed) curve exponentially approaches a limiting (full)
curve which coincides with the differential-momentum-scale-dependence
carried by the ultimate Wilson flow emerging from the Gaussian fixed point $%
P_{\text{G}}$ (lefthand full circle). The function $f_{1}(\overline{u}_{0})$
is numerically determined along this flow corresponding to the trajectory $%
T_{1}$ in ${\cal S}_{\text{c}}$. Notice that the convergence is numerically
well controlled independently of whether the initial point is chosen close
to $P_{\text{G}}$ or not. The righthand full circle indicate the infra-red
stable fixed-point-value $u_{0}^{*}=3.27039\cdots $. \label{fig4bis}

\item  Functional momentum-scale dependence displayed by $\overline{u}_{0}(l)
$ (projection onto the $u_{0}$-axis of ${\cal S}$) along definite Wilson's
flows on ${\cal S}_{\text{c}}$ (A). Open circles indicate the initial points
chosen in ${\cal S}_{\text{c}}$. Each full curve in (A) provides a
determination of the functional momentum-scale dependence. In order to
emphasize where the effective field theory (see section \ref{EFT})
practically occurs in each case, on figure (B) we have artificially
translated the ``time'' scales $l=-\ln (\lambda )$ (vertical dashed lines)
such that each actual Wilson's flow of (A) ``hits'' a given unique value of $%
\overline{u}_{0}$ at the same ``time''. The unique functional flow so
obtained illustrates the underlying unique differential flow of fig.(\ref
{fig4bis}).\label{fig04-04bis}

\item  Effective correction-exponent $\omega _{\text{eff}}(l)$ (Eq.(\ref
{omegaeff}) with $l=-\ln \lambda $) along $T_{1}$ (full curve) and along $%
T_{1}^{\prime }$ (dot-dashed curve). The estimate of the value of $\omega $
is $0.5955\pm 0.0025$.\label{fig11}

\item  The ``unusual'' infra-red stable one-dimensional submanifold in $%
{\cal S}_{\text{c}}$ ($T_{1}^{\prime }$) obtained from Eq.(\ref{4}) with $d=3
$ (projection onto the plane $(u_{0},v_{0})$). The full circle represents
the infra-red stable fixed point $f^{*}$ and the usual ``renormalized
trajectory'' $T_{1}$ emerging from $P_{\text{G}}$ is partially drawn (dashed
curve). Arrows indicate the infra-red direction. \label{fig15}

\item  The two infra-red attractive submanifolds in ${\cal S}_{\text{c}%
}^{(2)}$ $T_{2}$ and $T_{2}^{\prime }$ (full curves) of same degree of
stability but lower than on $T_{1}$ and on $T_{1}^{\prime }$ (dashed curves)
(projection onto the plane $(u_{0},v_{0})$). The initial values chosen are $%
r_{0\text{c}}(0)=\allowbreak -0.5832891\cdots $, $u_{0}(0)=\allowbreak
6.6615188\cdots $, $v_{0}(0)=\allowbreak 0$ ($T_{2}$) and $r_{0\text{c}%
}(0)=\allowbreak -0.42788517\allowbreak \cdots $, $u_{0}(0)=\allowbreak 2$, $%
v_{0}(0)=\allowbreak 26.13624$ ($T_{2}^{\prime }$). \label{fig16}

\item  Wilson's flows on the critical surface ${\cal S}_{\text{c}}$ obtained
from integration of Eq.(\ref{4}) at $d=4$ (projection onto the plane $%
(u_{0},v_{0})$). Open circles indicate the initial points chosen on the
canonical surface of ${\cal S}_{\text{c}}$: $(u_{0}(0),r_{0\text{c}%
}(0))=(2,\allowbreak -0.035806374\cdots )$; $(4,\allowbreak
-0.068682647\cdots )$; $(10,\allowbreak -0.1577796\cdots )$. Two flows
initialized at $(20,\allowbreak -0.2892822265\cdots )$ and $(40,\allowbreak
-0.52207031\cdots )$ are partially reproduced. The full circle represents
the infra-red stable fixed point $P_{\text{G}}$. Arrows indicate the
infra-red direction. The square roughly indicates where the effective field
theory could make sense for the Wilson trajectory initialized at $u_{0}(0)=10
$. \label{fig10}
\end{enumerate}


\begin{thebibliography}{99}
\bibitem{zin}  J. Zinn-Justin, {\em Euclidean Field Theory and Critical
Phenomena} (Oxford University Press, 1989), second edition (1993), third
edition (1996).

\bibitem{par2}  G. Parisi, in {\it Carg\`{e}se Summer School} (1973),
unpublished, and J. Stat. Phys. {\bf 23}, 49 (1980).

\bibitem{nic}  B. G. Nickel, in Phase Transitions Ed. by M. L\'{e}vy, J. C.
Le Guillou and J. Zinn-Justin (Plenum, New York and London, 1982).

\bibitem{nic2}  B. G. Nickel, Physica {\bf A177}, 189 (1991).

\bibitem{bakkin}  {G.A. Baker and J. M. Kincaid}, {Phys. Rev. Lett.} {\bf {42%
}}, {1431} (19{79}); {J. Stat. Phys.} {\bf {24}}, {469} (19{81}).

\bibitem{liufis}  A. Liu and M. E. Fisher, J. Stat. Phys. {\bf 58}, 431
(1990).

\bibitem{nic3}  {B.G. Nickel,} {Macromolecules} {\bf {24}}, {1358} (19{91}).

\bibitem{sokal}  A. D. Sokal, Europhys. Lett. {\bf 27}, 661 (1994).

\bibitem{3541}  B. Li, N. Madras and A. D. Sokal, J. Stat. Phys. {\bf 80},
661 (1995).

\bibitem{3532}  C. Bagnuls and C. Bervillier, Phys. Rev. Lett. {\bf 76},
4094 (1996); M. A. Anisimov, A. A. Povodyrev, V. D. Kulikov and J. V.
Sengers, Phys. Rev. Lett. {\bf 75}, 3146 (1995); {\em ibid.} {\bf 76}, 4095
(1996).

\bibitem{gellow}  M. Gell-Mann and F. E. Low, Phys. Rev. {\bf 95}, 1300
(1954).

\bibitem{gr}  E. Stueckelberg and A. Petermann, Helv. Phys. Acta {\bf 26},
499 (1953); L.V. Ovsyannikov, Dokl. Akad. Nauk SSSR, {\bf 109}, 1112 (1956),
in Russian; N. N. Bogoliubov and D. V. Shirkov, in ``{\em Introduction to
the Theory of Quantized Fields''} (Interscience, New York, 1959); C. G.
Callan, Phys.\ Rev. {\bf D2}, 1541 (1970); K. Symanzik, Comm. Math.\ Phys. 
{\bf 18}, 227 (1970); A. Petermann, Phys.\ Rep. {\bf 53}, 157 (1979).

\bibitem{brelegzin}  E. Br\'{e}zin, J. C. Le Guillou and J. Zinn-Justin, in
Phase Transitions and Critical Phenomena, Vol. VI, Ed. by C. Domb and M. S.
Green (Academic Press, New York, 1976).

\bibitem{wilkog}  K. G. Wilson and J. Kogut, Phys. Rep. {\bf 12C}, 77 (1974).

\bibitem{3480}  J. F. Nicoll, T. S. Chang and H. E. Stanley, Phys. Rev.
Lett. {\bf 33}, 540 (1974).

\bibitem{hashas}  A. Hasenfratz and P. Hasenfratz, Nucl.\ Phys. {\bf B270
[FS16]}, 687 (1986).

\bibitem{fel}  G. Felder, Comm. Math. Phys. {\bf 111}, 101 (1987).

\bibitem{weghou}  F. J. Wegner and A. Houghton, Phys. Rev. {\bf A8}, 401
(1973).

\bibitem{pol}  J. Polchinski, Nucl. Phys. {\bf B231}, 269 (1984).

\bibitem{3094}  J. Hughes and J. Liu, Nucl. Phys. {\bf B307}, 183 (1988).

\bibitem{luswei}  M. L\"{u}scher and P. Weisz, Nucl.\ Phys. {\bf B290 [FS20]}%
, 25 (1987).

\bibitem{LPA}  A. Margaritis, G. \'{O}dor and A. Patk\'{o}s, Z. Phys. {\bf %
C39}, 109 (1988); Yu. M. Ivanchenko, A. A. Lisyansky and A. E. Filippov, J.
Phys. {\bf A23}, 91 (1990); Phys. Lett. {\bf A150}, 100 (1990); A. E.
Filippov and S. A. Breus, Phys. Lett. {\bf A158}, 300 (1991); A. A.
Lisyansky, Yu. M. Ivanchenko and A. E. Filippov, J. Stat. Phys{\it .} {\bf 66%
}, 1667 (1992); A. E. Filippov and A. V. Radievskii, Sov. Phys-JETP {\bf 75}%
, 1022 (1992); Phys. Lett. {\bf A169}, 195 (1992); JETP Lett. {\bf 56}, 87
(1992); Yu. M. Ivanchenko and A. A. Lisyansky, Phys. Rev. {\bf A45}, 8525
(1992); S. A. Breus and A. E. Filippov, Physica {\bf A192}, 486 (1993); G.
Zumbach, Phys. Rev. Lett. {\bf 71}, 2421 (1993); Phys. Lett. {\bf A190}, 225
(1994); Nucl. Phys. {\bf B413}, 754 (1994); {\em ibid.} 771 (1994); P. E.
Haagensen, Y. Kubyshin, J. I. Latorre and E. Moreno, Phys. Lett{\it .} {\bf %
B323}, 330 (1994); T. R. Morris, Int. J. Mod. Phys{\it .} {\bf A8}, 2411
(1994); Phys. Lett. {\bf B329}, 241 (1994); Phys. Lett. {\bf B334}, 355
(1994); Nucl. Phys. {\bf B458 [FS]}, 477 (1996); A. E. Filippov, J. Stat.
Phys. {\bf 75}, 241 (1994); R. D. Ball, P. E. Haagensen, J. I. Latorre and
E. Moreno, Phys. Lett. {\bf B347}, 80 (1995); S.-B. Liao and J. Polonyi,
Phys. Rev. {\bf D51}, 4474 (1995); K. I. Aoki, K. Morikawa, W. Souma, J.-I.
Sumi and H. Terao, Prog. Theor. Phys. {\bf 95}, 409 (1996).

\bibitem{39}  C. Bagnuls and C. Bervillier, Saclay preprint T94/026 (1994),
unpublished.

\bibitem{bagber2}  C. Bagnuls and C. Bervillier, Phys. Rev. Lett. {\bf 60},
1464 (1988).

\bibitem{bagber8}  C. Bagnuls and C. Bervillier, Phys. Rev. {\bf B41}, 402
(1990).

\bibitem{inprep2}  C. Bagnuls and C. Bervillier, Phys. Lett{\it .} {\bf A195}%
, 163 (1994).

\bibitem{dimreg}  G. 't Hooft and M. Veltman, Nucl. Phys. {\bf B44}, 189
(1972); J. F. Ashmore, Lett. Nuovo Cimento {\bf 4}, 289 (1972); C. G.
Bollini and J. J. Giambiagi, Phys. Lett. {\bf 40B}, 566 (1972); G. 't Hooft,
Nucl. Phys. {\bf B61}, 455 (1973); P. Butera, G. M. Cicuta and E. Montaldi,
in Renormalization and Invariance in QFT, p. 1, {\em Ed. by} E. R. Caianello
(Plenum Press, N.Y. and London, 1974).

\bibitem{wei}  S. Weinberg, Phys. Rev. {\bf D8}, 3497 (1973).

\bibitem{dys1}  {F. Dyson}, {Phys. Rev.} {\bf {85}}, {\ 631} (19{52}); C. S.{%
\ Lam}, {Nuovo Cimento} {\bf {55}}, {258} (19{68}); L. N. Lipatov, Zh. Eksp.
Teor. Fiz. {\bf 72}, 411 (1977) [Sov. Phys.-JETP {\bf 45}, 216 (1977)]; Zh.
Eksp. Teor. Fiz. {\bf 25}, 116 (1977) [JETP Lett. {\bf 25}, 104 (1977)].

\bibitem{eckmagsen}  J. P. Eckmann, J. Magnen and R. S\'{e}n\'{e}or, Comm.
Math. Phys., {\bf 39}, 251 (1975).

\bibitem{439}  K. G. Wilson and M. E. Fisher, Phys. Rev. Lett. {\bf 28}, 240
(1972).

\bibitem{sym}  {K. Symanzik}, {Nuovo Cimento Lett.} {\bf {8}}, {\ 771} (19{73%
}); G. Parisi, Nucl. Phys. {\bf B150}, 163 (1979); {R. Jackiw and S.
Templeton}, {Phys. Rev.} {\bf {D23}}, {\ 2291} (19{81}); M. C.{\ Berg\`{e}re
and F. David}, {Ann. Phys. (N.Y.)} {\bf {142}}, {\ 416} (19{82}).

\bibitem{gro}  D. J. Gross, in Methods in Field Theory Ed. by R. Balian and
J. Zinn-Justin (North Holland, Amsterdam, 1976); D. J. Amit, Field Theory, 
{\em The Renormalization Group and Critical Phenomena} (World Scientific,
Singapore, 1984); G. Parisi, {\em Statistical Field Theory} (Addison-Wesley
Publ., 1988); M. Le Bellac, {\em Des Ph\'{e}nom\`{e}nes Critiques aux Champs
de Jauge} (Inter Editions/Editions du CNRS, 1988). C. Itzykson and J. M.
Drouffe, {\em Statistical Field Theory} Vol. I (Cambridge University Press,
Cambridge, 1989).

\bibitem{2876}  B. Kr\"{u}ger and L. Sch\"{a}fer, J. Phys. I (France) {\bf 4}%
, 757 (1994). L. Sch\"{a}fer, Phys. Rev{\it .} {\bf E50}, 3517 (1994).

\bibitem{gro1}  D. J. Gross, in Recent Developments in Quantum Field Theory
Ed. by J. Ambjorn, {B. J. Durhuus and J. L. Petersen (}{Elsevier Science
Publishers B. V., 19}{85).}

\bibitem{shi}  D. V. Shirkov, Int. J. Mod. Phys. {\bf A3}, 1321 (1988).

\bibitem{3477}  J. F. Nicoll, T. S. Chang and H. E. Stanley, Phys. Rev. {\bf %
B12}, 458 (1975).

\bibitem{wil11}  K. G. Wilson, Adv. Math. {\bf 16}, 444 (1975).

\bibitem{dys}  F. Dyson, Comm. Math. Phys. {\bf 12}, 91 (1969).

\bibitem{sym2}  K. Symanzik, in {New Developments in Quantum Field Theory
and Statistical Mechanics Ed. by }{M. L\'{e}vy and P. K. Mitter (}{Plenum,
New York, 19}{77); }in Recent Development in Gauge Field Theories Ed. by G
't Hooft et al (Plenum, New-York, 1980); in {Mathematical Problems in
Theoretical Physics. Lecture Notes in Physics Ed. by }{R. Schrader et al (}{%
Springer, Berlin, 19}{82); Nucl. Phys. }{\bf B226}{, 205 (19}{83); }Nucl.
Phys. {\bf B226}{, 187 (19}{83).}

\bibitem{3344}  A. E. Filippov, {\it JETP Lett.} {\bf 60}, 141 (1994).

\bibitem{weg}  F. J. Wegner, Phys.\ Rev. {\bf B5}, 4529 (1972); in {Phase
Transitions and Critical Phenomena Vol. }{VI Ed. by }{C. Domb and M. S.
Green (}{Academic Press, New York, 19}{76).}

\bibitem{geo}  H. M. Georgi, in The New Physics Ed. by P. Davies (Cambridge
University Press, 1989); G. P. Lepage, in From Actions to Answers, proc. of
the 1989 TASI Summer School, Ed. by T. De Grand and D. Toussaint (World
Scientific, Singapore, 1990).

\bibitem{hasnag}  P. Hasenfratz and J. Nager, Z. Phys. {\bf C37}, 477 (1988).

\bibitem{gronev}  D.J. Gross and A. Neveu, Phys. Rev. {\bf D10}, 3235 (1974).

\bibitem{rimwei}  C. Rim and W. I. Weisberger, {\bf D32}, 3244 (1985).

\bibitem{par7}  G. Parisi, Phys.\ Lett. {\bf 76B}, 65 (1978); Phys. Rep. 
{\bf 49}, 215 (1979).

\bibitem{berdav1}  M. C. Berg\`{e}re and F. David, Phys.\ Lett. {\bf 135B},
253 (1984).

\bibitem{felmagrivsen}  J.S. Feldman, J. Magnen, V. Rivasseau and R.
S\'{e}n\'{e}or, Comm. Math. Phys. 109, 437 (1987); K. Gawedski and A.
Kupiainen, Phys. Rev. Lett. 54, 92 (1985); Comm. Math. Phys. 99, 197 (1985).

\bibitem{bak5}  G. A. Baker, Jr, Phys. Rev. {\bf B15}, 1552 (1977).

\bibitem{legzin}  J. C. Le Guillou and J. Zinn-Justin, Phys. Rev. Lett. {\bf %
39}, 95 (1977); Phys. Rev. {\bf B21}, 3976 (1980); J. Phys. Lett. (France) 
{\bf 46}, L137 (1985).

\bibitem{golrie}  G. R. Golner and E. K. Riedel, Phys. Lett. {\bf A58}, 1365
(1976).

\bibitem{newrie}  K. E. Newman and E. K. Riedel, Phys. Rev. {\bf B30}, 6615
(1984).

\bibitem{zin7}  J. Zinn-Justin, in Phase Transitions, Ed. by M. L\'{e}vy, J.
C. Le Guillou and J. Zinn-Justin (Plenum, New York and London, 1982).
\end{thebibliography}
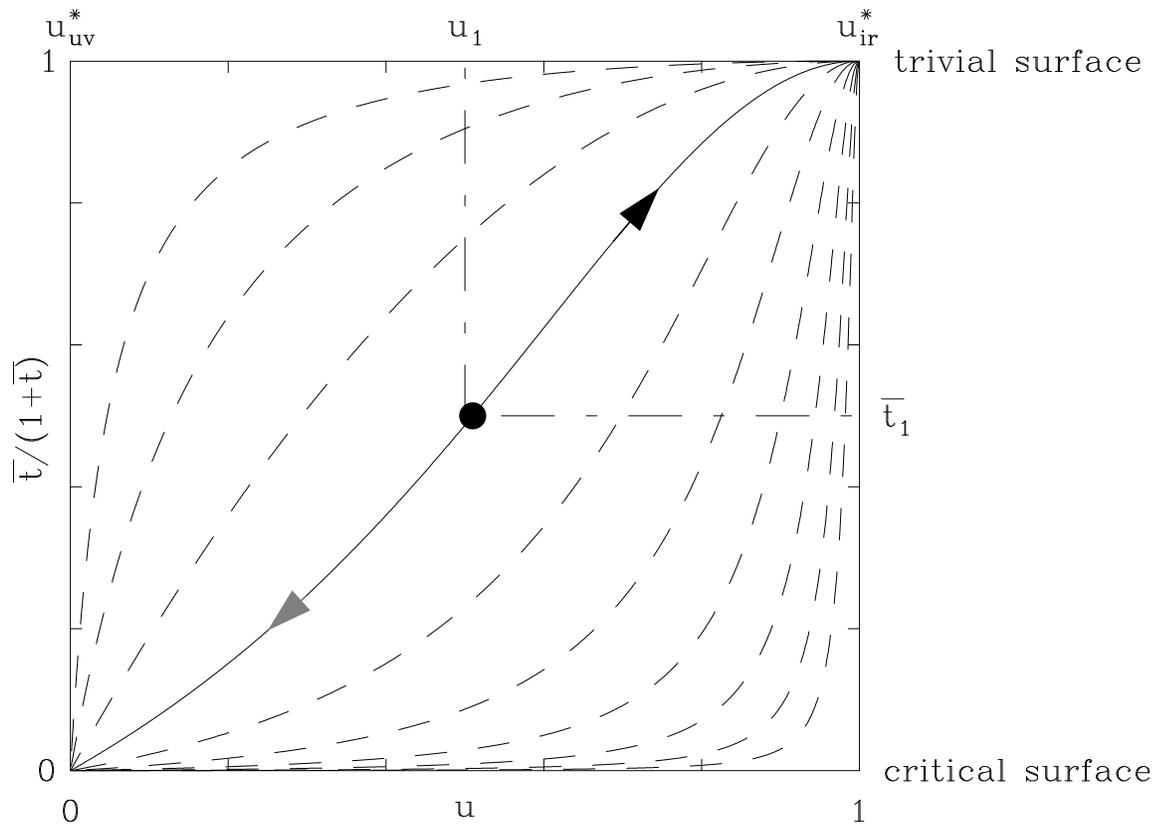
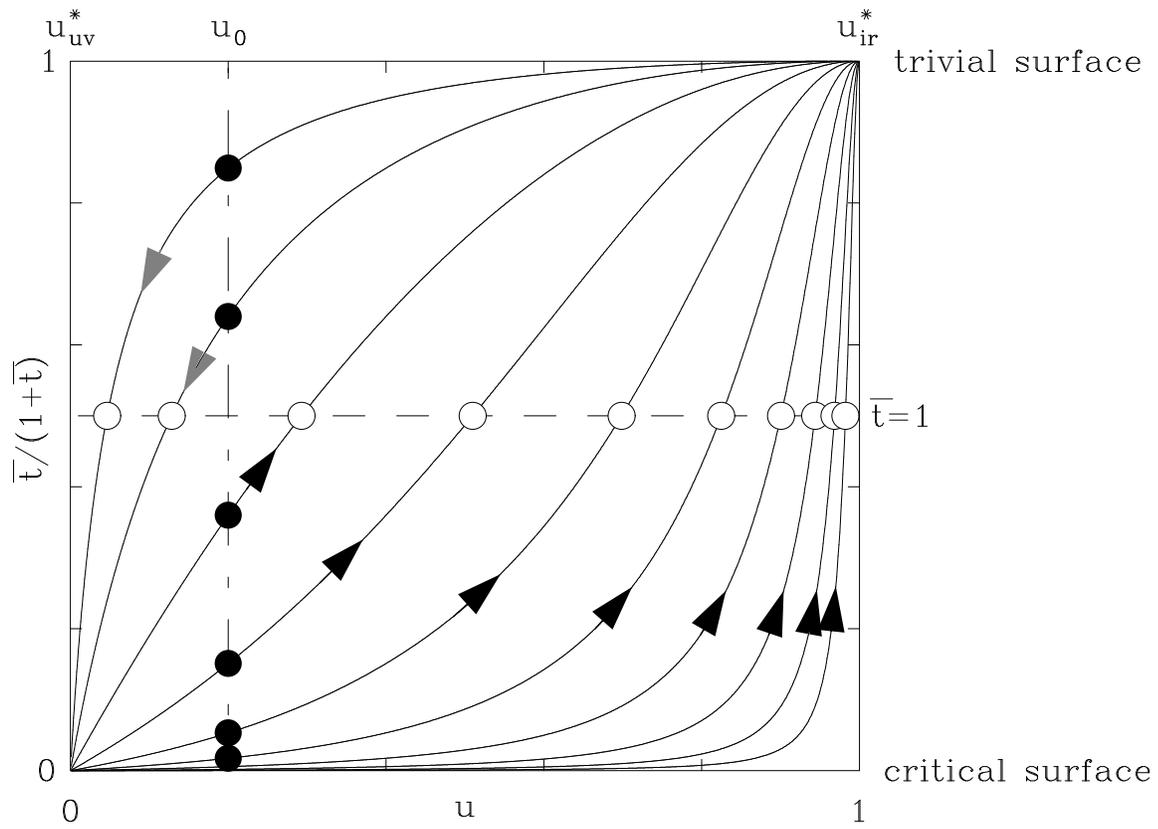
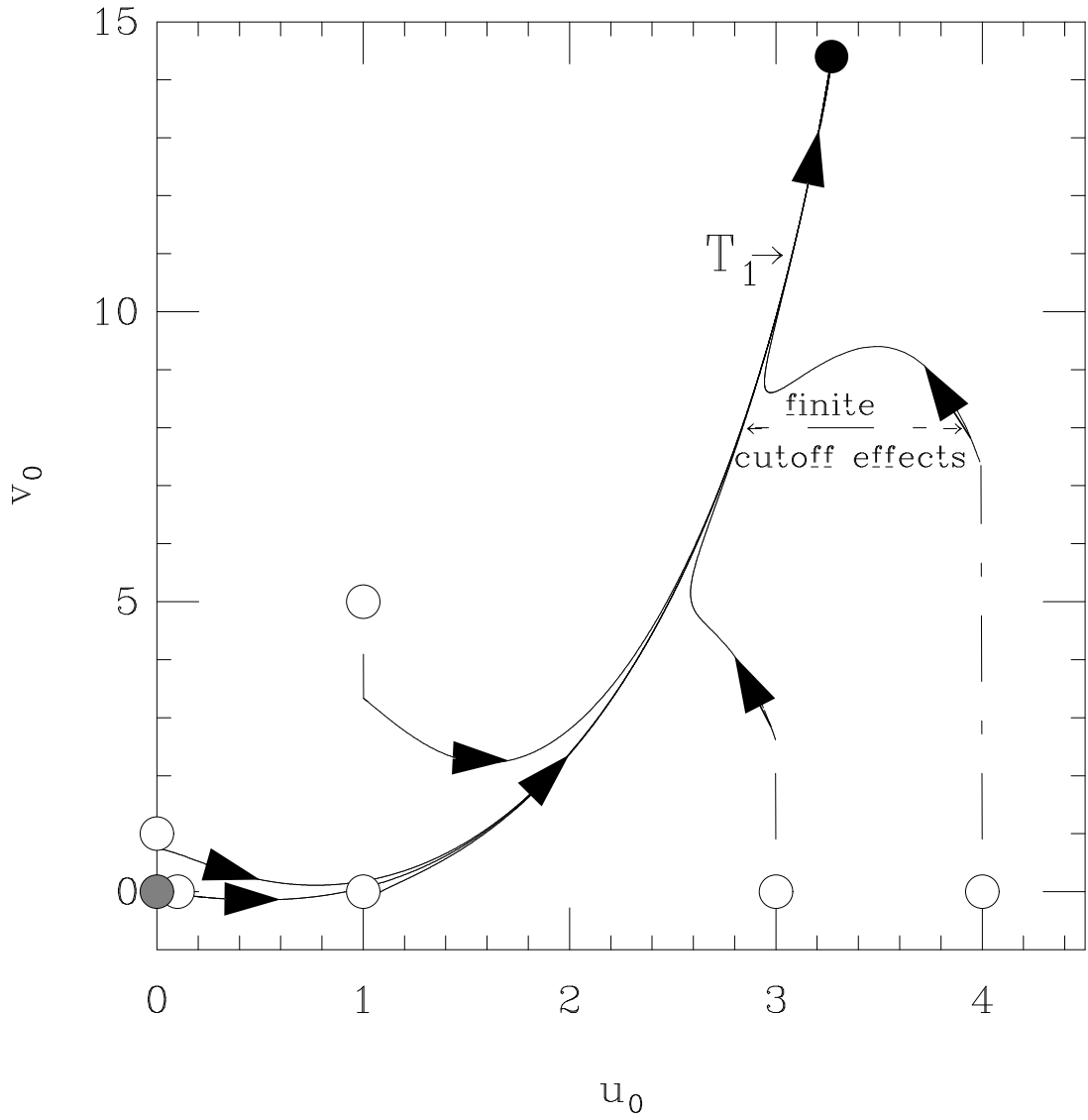
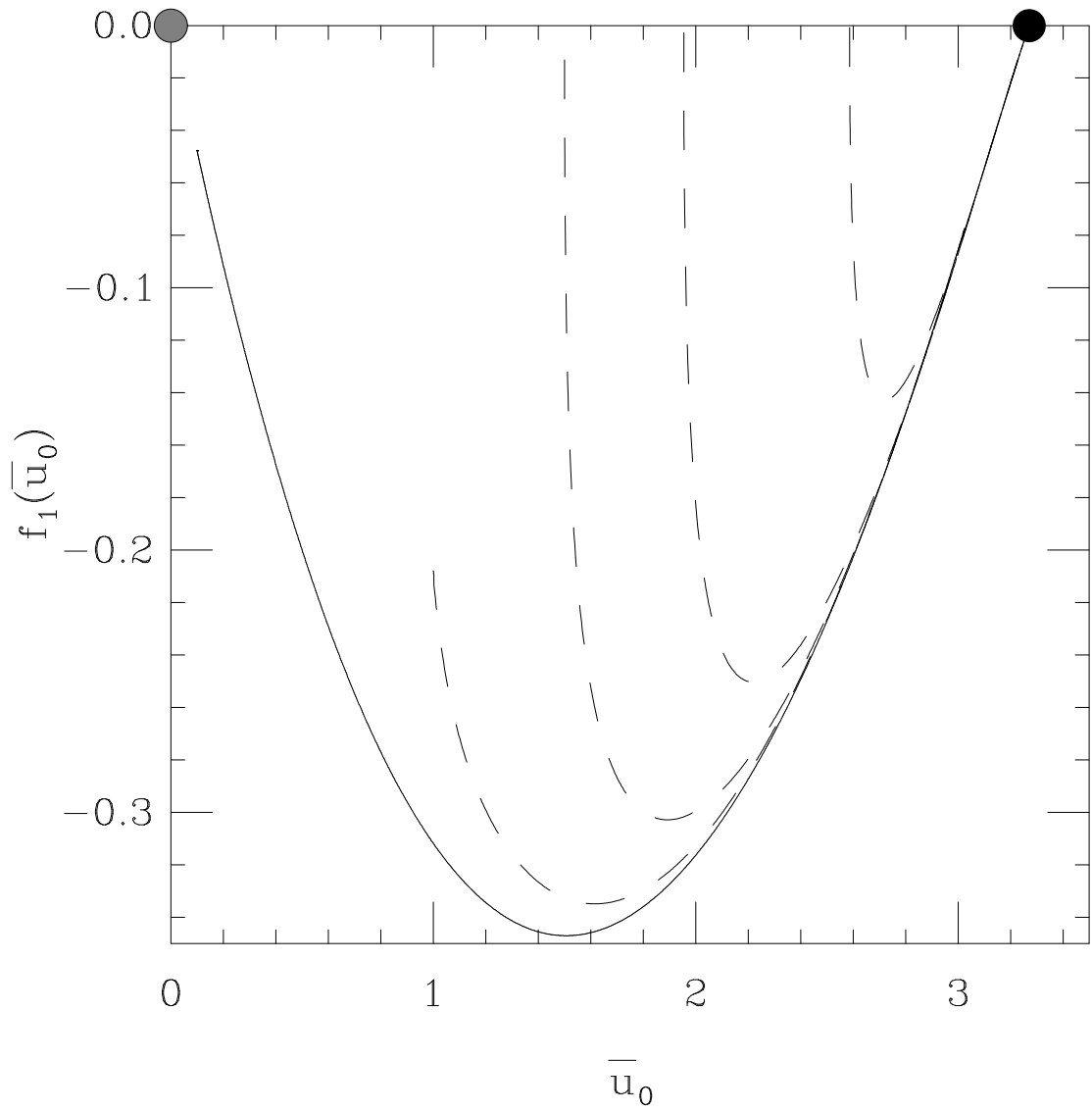
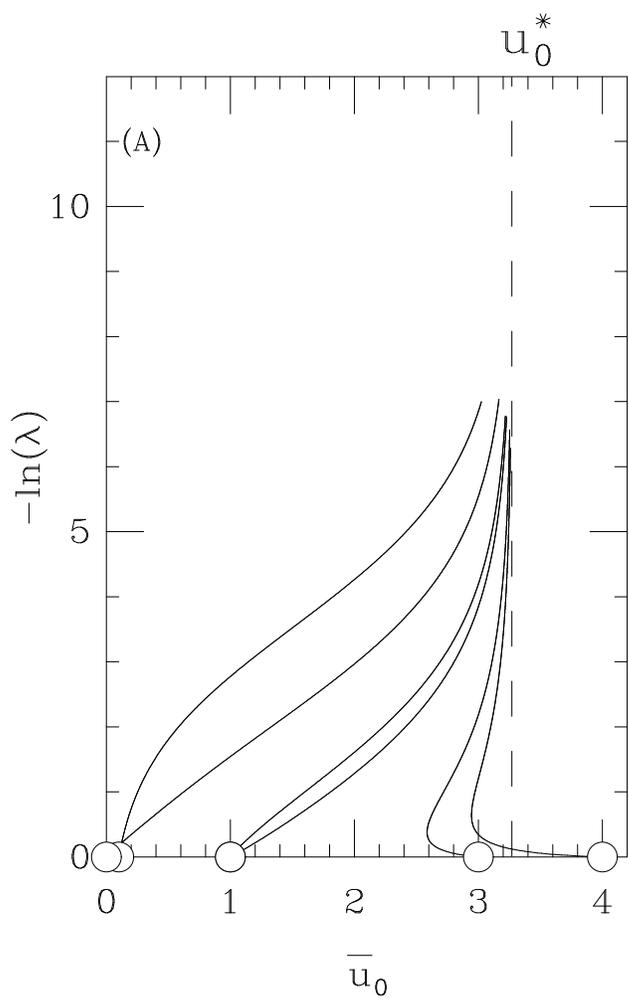
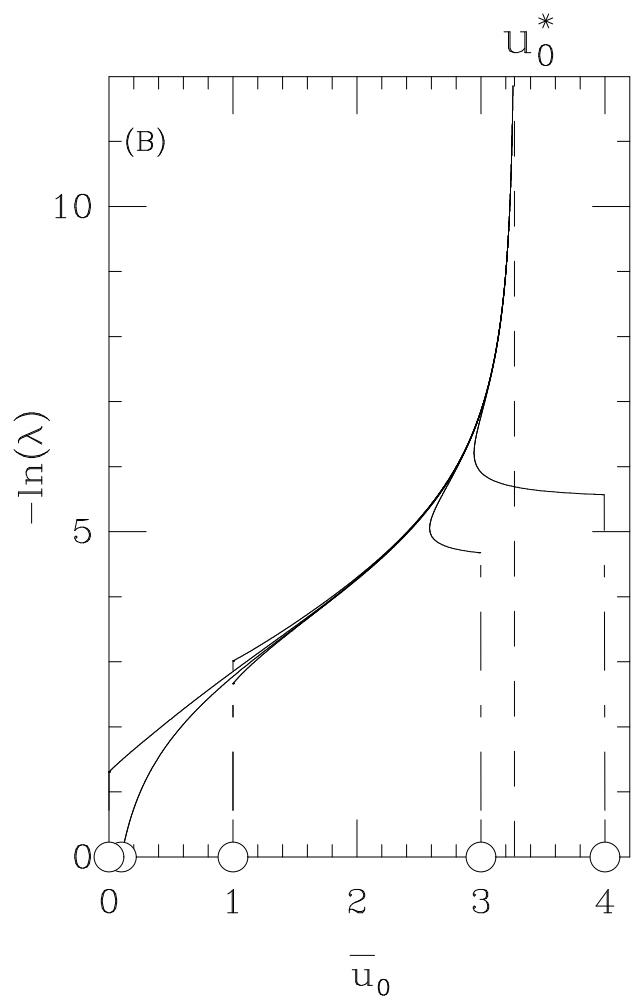
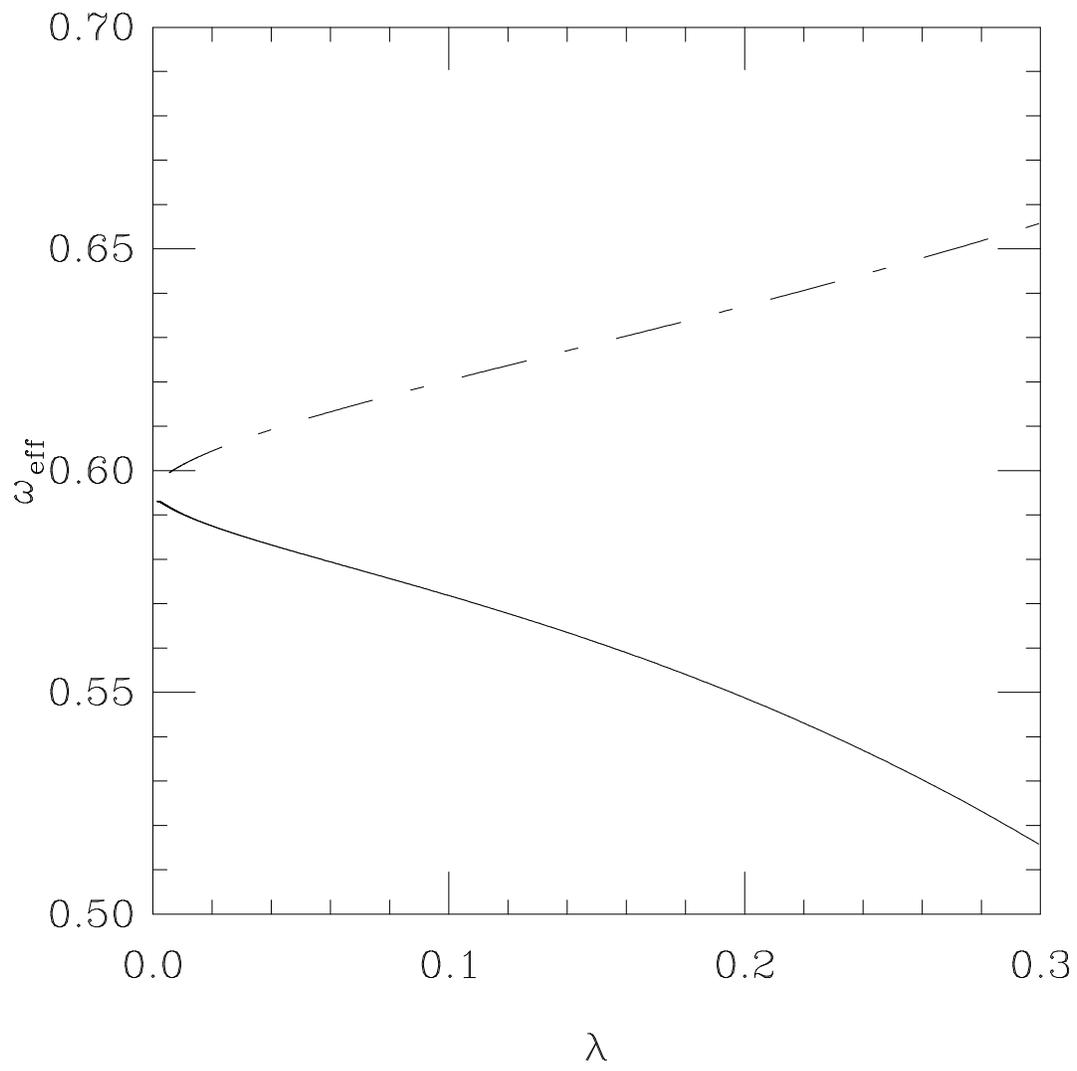
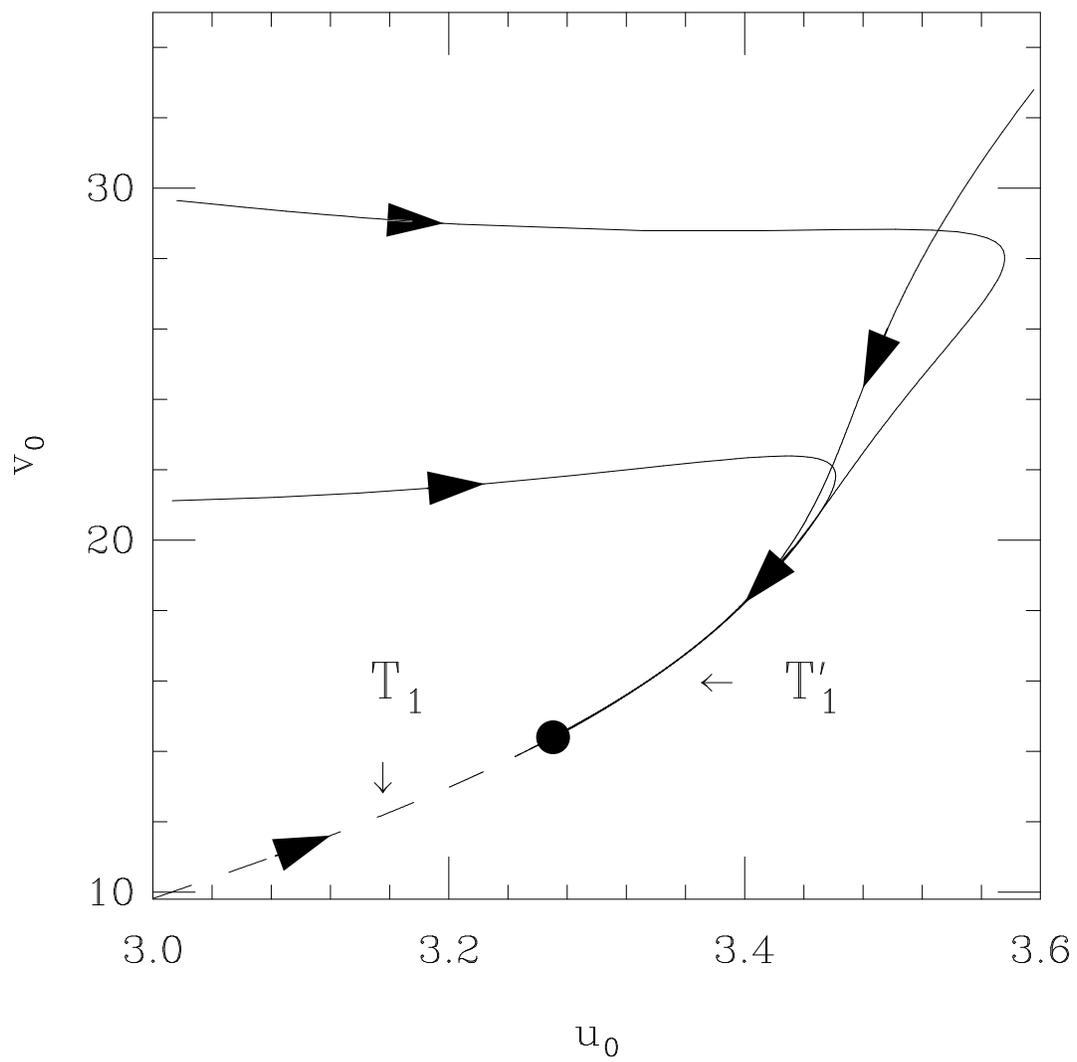
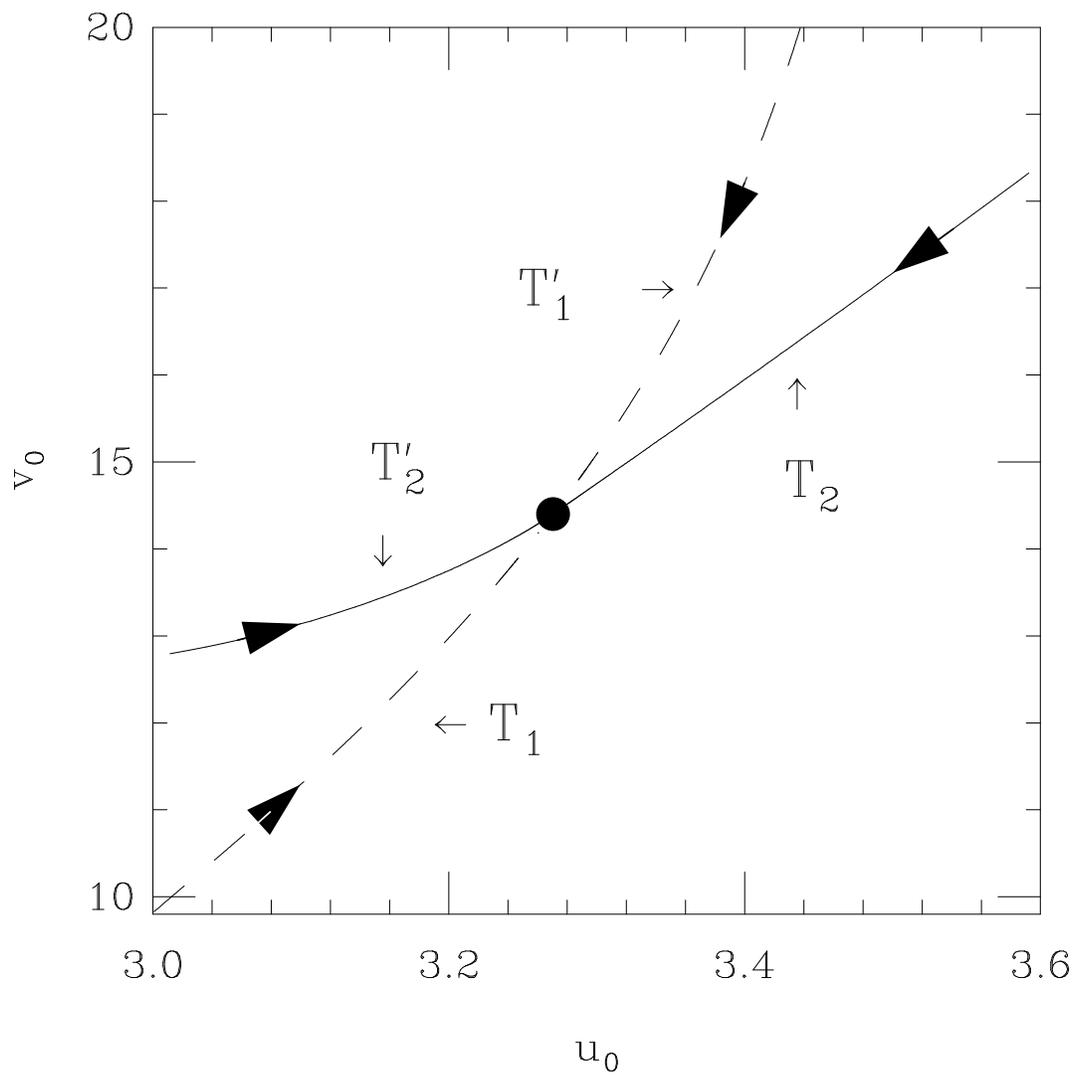
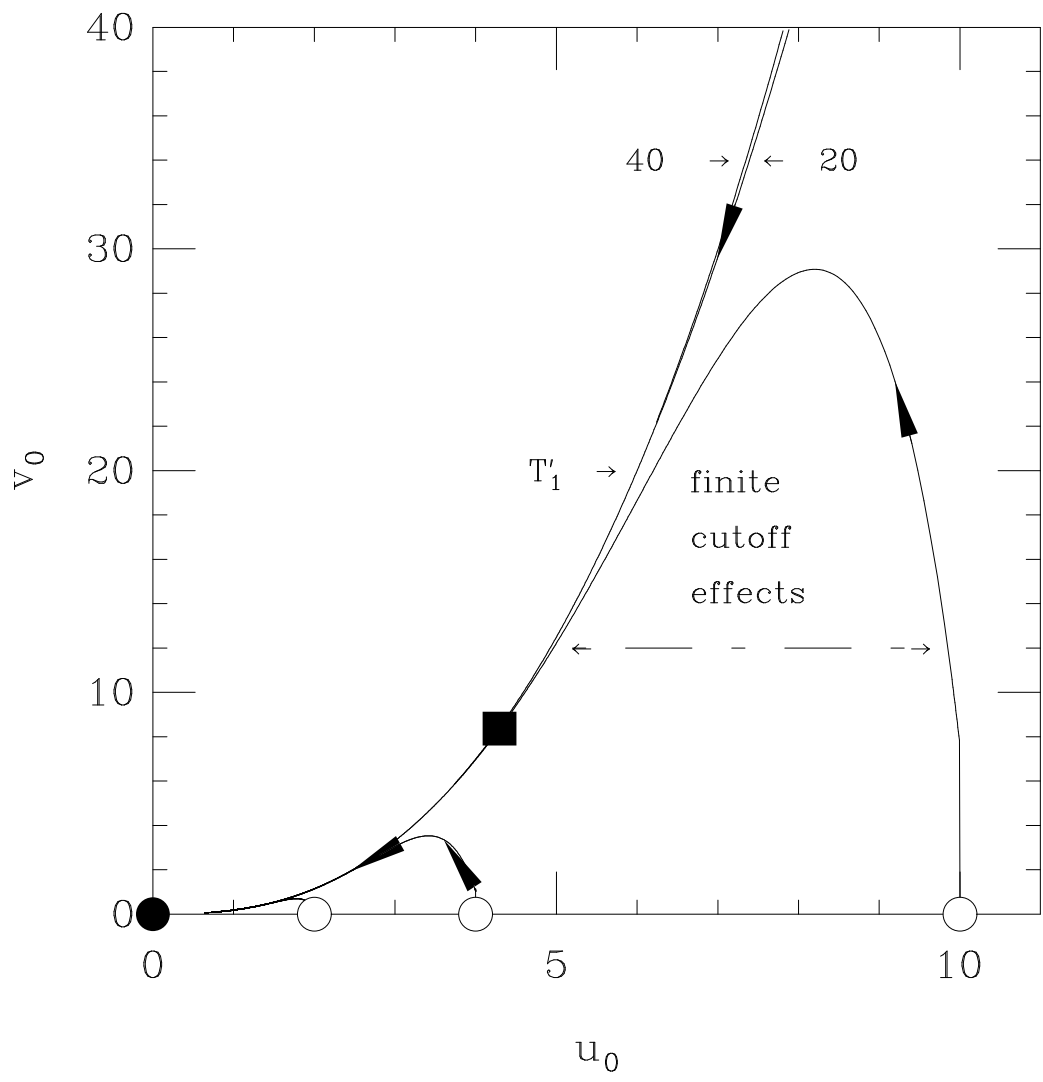
\end{document}